%% file: sample-manuscript.tex
\documentclass[acmlarge,screen]{acmart}

\AtBeginDocument{%
  }

\usepackage{siunitx}
\usepackage{subfloat}
\input{scifilab}

\usepackage{glossaries}
\usepackage{multirow}
\usepackage{pifont}

\begin{document}

\title{SonicID: User Identification on Smart Glasses with Acoustic Sensing}

\author{Ke Li}
\affiliation{%
  \institution{Cornell University}
  \city{Ithaca}
  \country{USA}}
\email{kl975@cornell.edu}
\orcid{0000-0002-4208-7904}

\author{Devansh Agarwal}
\affiliation{%
  \institution{Cornell University}
  \city{Ithaca}
  \country{USA}}
\email{da398@cornell.edu}
\orcid{0009-0005-1338-9275}

\author{Ruidong Zhang}
\affiliation{%
  \institution{Cornell University}
  \city{Ithaca}
  \country{USA}}
\email{rz379@cornell.edu}
\orcid{0000-0001-8329-0522}

\author{Vipin Gunda}
\affiliation{%
  \institution{Cornell University}
  \city{Ithaca}
  \country{USA}}
\email{vg245@cornell.edu}
\orcid{0009-0000-5500-2183}

\author{Tianjun Mo}
\affiliation{%
  \institution{Cornell University}
  \city{Ithaca}
  \country{USA}}
\email{tm659@cornell.edu}
\orcid{0009-0003-8405-4769}

\author{Saif Mahmud}
\affiliation{%
  \institution{Cornell University}
  \city{Ithaca}
  \country{USA}}
\email{sm2446@cornell.edu}
\orcid{0000-0002-5283-0765}

\author{Boao Chen}
\affiliation{%
  \institution{Cornell University}
  \city{Ithaca}
  \country{USA}}
\email{bc526@cornell.edu}
\orcid{0000-0002-3527-9481}

\author{François Guimbretière}
\affiliation{%
  \institution{Cornell University}
  \city{Ithaca}
  \country{USA}}
\email{fvg3@cornell.edu}
\orcid{0000-0002-5510-6799}

\author{Cheng Zhang}
\affiliation{%
  \institution{Cornell University}
  \city{Ithaca}
  \country{USA}}
\email{chengzhang@cornell.edu}
\orcid{0000-0002-5079-5927}
\renewcommand{\shortauthors}{Li et al.}

\input{0_abstract}


\setcopyright{acmlicensed}
\acmJournal{IMWUT}
\acmYear{2024} \acmVolume{8} \acmNumber{4} \acmArticle{169} \acmMonth{12}\acmDOI{10.1145/3699734}

\maketitle

\input{1_intro}
\input{2_related_work}
\input{4_system_design}
\input{5_form_factor}
\input{6_user_study}
\input{7_results}
\input{8_discussion}
\input{10_conclusion}
\input{11_acks}

\bibliographystyle{ACM-Reference-Format}
\bibliography{sample-base}

\input{12_appendix}

\end{document}

%% file: scifilab.tex
\usepackage{color,soul}
\usepackage{booktabs} 
\usepackage{subfig} 
\usepackage{graphicx}

\usepackage[utf8]{inputenc}
\usepackage[T1]{fontenc}

\usepackage{siunitx}
\newcommand{\hide}[1]{}

\sethlcolor{yellow}
\ifdefined\scifihidecomments
    \newcommand{\cheng}[1] {}
    \newcommand{\hyunc}[1] {} 
    \newcommand{\yax}[1] {} 
    \newcommand{\tuoc}[1] {} 
    \newcommand{\sony}[1] {} 
    \newcommand{\ReviewerFeedback}[1] {}  
    \newcommand{\fran}[1] {}
    \newcommand{\rz}[1]{}
    \newcommand{\lw}[1] {}
    \newcommand{\mose}[1] {} 
    \newcommand{\ke}[1] {} 

\else
    \definecolor{burntorange}{rgb}{0.8, 0.33, 0.0}
    \definecolor{cadmiumgreen}{rgb}{0.0, 0.42, 0.24}
    \definecolor{cobalt}{rgb}{0.0, 0.28, 0.67}
    \definecolor{amber}{rgb}{1.0, 0.75, 0.0}
    \definecolor{fashionfuchsia}{rgb}{0.96, 0.0, 0.63}
    \definecolor{brightcerulean}{rgb}{0.11, 0.67, 0.84}
    \definecolor{frenchblue}{rgb}{0.0, 0.45, 0.73}
    \definecolor{darkslateblue}{rgb}{0.28, 0.24, 0.55}
    \definecolor{cerulean}{rgb}{0.0, 0.48, 0.65}
    \definecolor{darkpastelgreen}{rgb}{0.01, 0.75, 0.24}

    \newcommand{\hyunc}[1] { \textcolor{burntorange}{[{\hl{hyunc:}} {#1}}]}
    \newcommand{\yax}[1] { \textcolor{magenta}{[{\hl{yaxuan:}} {#1}]}}
    \newcommand{\tuoc}[1] { \textcolor{darkpastelgreen}{[{\hl{tuochao:}} {#1}]}}
    \newcommand{\sony}[1] { \textcolor{blue}{[{\hl{songyun:}} {#1}]}}
    \newcommand{\ReviewerFeedback}[1] { \textcolor{brightcerulean}{[{Reviewer Feedback:} {#1}}]}
    \newcommand{\fran}[1]{\textcolor{burntorange}{[{francois:}{#1}}]}
    \newcommand{\rz}[1]{\textcolor{teal}{[{Ruidong: }{#1}]}}
    \newcommand{\lw}[1]{\textcolor{fashionfuchsia}{[{liuwei:}{#1}}]}
    \newcommand{\mose}[1]{\textcolor{burntorange}{[{mose:}{#1}}]}
    \newcommand{\ke}[1] { \textcolor{red!55!yellow}{[{Ke:} {#1}}]}

\fi

\ifdefined\scifiblind
    \newcommand{\blind}[1]{[omitted for blind review]}
\else
    \newcommand{\blind}[1]{#1} 
\fi

%% file: 0_abstract.tex
\begin{abstract}
Smart glasses have become more prevalent as they provide an increasing number of applications for users. They store various types of private information or can access it via connections established with other devices. Therefore, there is a growing need for user identification on smart glasses. In this paper, we introduce a low-power and minimally-obtrusive system called SonicID, designed to authenticate users on glasses. SonicID extracts unique biometric information from users by scanning their faces with ultrasonic waves and utilizes this information to distinguish between different users, powered by a customized binary classifier with the ResNet-18 architecture. SonicID can authenticate users by scanning their face for 0.06 seconds. A user study involving 40 participants confirms that SonicID achieves a true positive rate of 97.4\%, a false positive rate of 4.3\%, and a balanced accuracy of 96.6\% using just 1 minute of training data collected for each new user. This performance is relatively consistent across different remounting sessions and days. Given this promising performance, we further discuss the potential applications of SonicID and methods to improve its performance in the future.

\end{abstract}

\keywords{User Identification, Smart Glasses, Acoustic Sensing, Machine Learning}

%% file: 1_intro.tex
\section{Introduction}

Smart glasses, such as Vuzix Smart Glasses~\cite{vuzix}, Lenovo ThinkReality A3~\cite{ThinkReality}, Meta Ray-Ban~\cite{ray-ban}, Snap Spectacles 3~\cite{snap}, etc., have been growing in popularity and providing users with more and more interactive applications in recent years. Similar to smartphones and other wearable devices (e.g. smart watches and VR headsets), smart glasses start to store users' personal data and private information including their private messages, photos, videos and banks accounts. Leakage of private information and inconvenience might be induced to certain users if bad actors obtain this information from their smart glasses. Moreover, some smart glasses are connected to other devices such as smartphones and smart TVs. It is possible for smart glasses without user authentication to be utilized by bad actors to give unwanted commands to or display inappropriate contents on these connected devices. Therefore, there is an increasing need for implementing user authentication methods on smart glasses. 

However, it is difficult to directly apply traditional frontal-camera-based authentication techniques, e.g. FaceID, on glasses because they require placing a camera in front of the user to capture their face. Researchers have put efforts into deploying cameras on glasses to capture biometric information of the user from their facial contours~\cite{lim2023cauth}, eye movements~\cite{zhang2018continuous}, or iris features~\cite{Yunghui2017Iris}. Though these camera-based methods achieve satisfactory performance of user authentication on glasses, they require the placement of a camera which is relatively large compared to the small size of glasses and might partially block the user's view. Furthermore, the cameras on commodity smart glasses, though can be smaller in size, usually point forwards to capture the environmental information. Barriers exist if they are to be directly used to capture the biometric information of the user's face. Password-based authentication methods have been explored on smart glasses as well, especially with voice input \cite{bailey2014typing, Yadav2015VBP, Li2021Gtalker}. In addition, behavioral biometrics, such as tapping gestures \cite{Rasel2018GlassPass}, touch gestures \cite{Jagmohan2016Gesture,Ge2017GlassGuard}, and voice commands \cite{Ge2017GlassGuard}, have been leveraged to authenticate users. These methods require the user's interaction and are notably vulnerable to shoulder-surfing attacks since these gestures are entered on a touchpad and the vocalized commands can be overheard by bystanders.

In this paper, we present SonicID, a low-power and minimally-obtrusive solution to user authentication on smart glasses, which identifies users automatically as soon as they put on the glasses, using active acoustic sensing. We place one pair of speaker and microphone on each side of a pair of glasses. The sensors are instrumented on the hinges of the glasses, pointing downwards at the user's face. The core sensing principle of SonicID is to \textit{use a sonar-like technique to scan the shape of the user's face as the biometric information for authentication}. Similar to how FaceID scans the user's face with infrared cameras, when a user wears the glasses instrumented with SonicID, it scans the user's face automatically with acoustic signals, and then compares scanned patterns with the profiles of registered users to determine whether or not this user can have access to the device. 

Specifically, the speakers on the glasses emit encoded inaudible Frequency Modulated Continuous Waves (FMCW) towards the user's face after they put on the glasses. The signals continuously and repeatedly sweep at different frequency ranges on two sides to avoid interference. The transmitted signals are reflected by the user's face and captured by the microphones on the glasses. After signal processing using the received signals and transmitted signals, SonicID obtains unique acoustic patterns related to this user's face. Considering that everyone has different facial shapes, the user's biometric information is included in the captured acoustic patterns which can be learned to distinguish this user from other users and thus authenticate them securely. In SonicID, we feed the processed acoustic patterns into a customized binary classifier using the ResNet-18 architecture to extract distinct features so as to authenticate users. The classifier is trained with the training data collected when a user registers on the smart glasses and it keeps authenticating users in later usage of the device just like traditional user authentication techniques.

To validate the performance of our SonicID system, we conducted a user study with 40 participants. The study results demonstrate that a new user needs to provide 1 minute of data to train the binary classifier on the first day of using the smart glasses, to achieve a true positive rate of 97.4\%, a false positive rate of 4.3\%, and a balanced accuracy of 96.6\%. In this study, we also asked all participants to come back to test the performance of the system on two other different days. The results show that the performance of the system remains relatively consistent across three different days, especially for the false positive rate. If 15 seconds of data is collected on a second day to fine-tune the trained model, the authentication performance of SonicID further improves across days, reaching a balanced accuracy around 90\% on Day 2 and Day 3. This study validates that SonicID achieves a satisfactory and stable authentication performance on smart glasses. Compared with frontal-camera-based authentication techniques such as FaceID, SonicID does not need a camera to be placed in front of the user. Other traditional authentication methods which take fingerprints or passwords as input can eliminate the need for a frontal camera. However, they require users' operations for authentication while SonicID automatically logs valid users in when they put the device on. Moreover, SonicID is capable of successfully authenticating users by scanning their face for only 0.06 seconds. Each time it authenticates users, SonicID, excluding the machine learning model inference part, consumes energy as little as 32.7 mJ. This low-power feature guarantees that SonicID can work approximately 31,500 times theoretically with the battery of Meta Ray-Ban (154 mAh) if the model inference of SonicID is implemented in real-time on a low-power chip such as MAX78002~\cite{MAX78002} in the future work and no other operations are running on the device.

The main contributions of this paper are summarized as follows:

\begin{itemize}
    \item We proposed a low-power and minimally-obtrusive solution to the user authentication task on smart glasses, using active acoustic sensing. It demonstrated that the shape of the user's face can be scanned with acoustic signals as biometric information for authentication.
    \item We prototyped the system with one pair of speaker and microphone on each side of the glasses. The system maintains low-power, consuming only 32.7 mJ of energy every time it authenticates the user, excluding the machine learning model inference part.
    \item We conducted a user study with 40 participants, validating that only 1 minute of training data for each new user guarantees that the system achieves a true positive rate of 97.4\%, a false positive rate of 4.3\%, and a balanced accuracy of 96.6\%. The performance remains relatively consistent in evaluations across different days.
\end{itemize}

%% file: 2_related_work.tex
\section{Related Work}
\label{Sec: Related Work}
In this section, we present the related work of SonicID in the following three scopes: (1) Authentication methods on non-wearables; (2) Authentication methods on wearable devices; (3) Comparison between our work and other authentication methods on wearables.

\subsection{Authentication on Non-wearable Devices}
Authentication mechanisms are critical in devices like smartphones, tablets, and computers that store sensitive personal information, provide access to banking services, and control security systems \cite{Karlson2009BorrowPhone}. Traditional authentication methods, such as passwords and Personal Identification Numbers (PINs), are prevalent but not without shortcomings. Users may forget these credentials, and they are susceptible to security breaches, including shoulder-surfing attacks \cite{Tari2006Shoulder}. An alternative approach, pattern-based passwords, leverages the pictorial superiority effect by allowing users to remember shapes, offering a more intuitive recall process \cite{Jermyn1999GraphicPasswords, Nelson1976Pictorial, Standing1973Pictures}. However, these patterns are vulnerable to both shoulder-surfing and smudge attacks \cite{Adam2010Smudge}.
        
Biometric authentication has gained popularity due to its perceived security and ease of use. Fingerprint sensing is widely used in current devices \cite{AppleTouchId}. Face recognition technologies, although increasingly common, can be compromised using photographs or videos of the user \cite{duc2009your}. To counter this, systems like Apple's FaceID employ 3D reconstruction of facial features \cite{AppleFaceId}. Iris scans offer another biometric alternative, noted for their uniqueness and difficulty to replicate \cite{John1991Iris}. Additionally, multimodal biometric systems have been explored, combining different physical attributes for enhanced security. For instance, Rokita et al. \cite{Rokita2008FaceAndhand} integrates face and hand images, while Marcel et al. \cite{Marcel2010Mobio} combines face and speech recognition. EchoPrint \cite{Zhou2018EchoPrint} merges facial recognition with acoustic analysis to counter spoofing attempts. VoiceLive \cite{ZhangVoiceLive2016} and VoiceGesture \cite{Zhang2017ArticulatoryGesture} utilize unique vocal characteristics and articulatory gestures, respectively, to thwart replay attacks. AirSign \cite{Shao2020AirSignSA} innovatively employs acoustic sensing and motion sensors for air signature-based authentication.
        
Behavioral biometrics also provide promising avenues for authentication. These methods analyze user-specific behaviors, such as pattern input dynamics \cite{Luca2012Touch}, activity patterns \cite{EhatishamulHaq2017AuthenticationOS, Malik2019ADLAuth}, gait \cite{zou2020deep}, and even unique hand movements and orientations \cite{Sitová2016HMOG}. BreathPrint \cite{Chauhan2017BreathPrintBA} further extends this domain by using audio features from individual breathing patterns. Such techniques offer continuous and passive authentication, enhancing both security and user convenience. 

However, these mature authentication technologies above usually cannot be easily transplanted to wearable devices considering their small size and limited battery capacity.

\subsection{Authentication on Wearable Devices}
In recent years, the proliferation of wearable devices, such as smartwatches, smart rings, earphones, and smart glasses, has been notable. These devices, as highlighted by \cite{Ometov2021ASO}, are increasingly integrating into our daily lives, storing sensitive data ranging from SMS messages \cite{Do2017IsTD} to health information \cite{Abdelhamid2021FitnessTI}. Moreover, their application in critical functions, such as contactless payments \cite{Imad2022Payments} and authorizing access to laptops \cite{Shazad2017IOT}, underscores the necessity of robust access control systems. Consequently, there has been a significant thrust in research towards developing effective authentication methods for these devices.

\subsubsection{Smart Glasses}
To enhance the security of smart glasses, a variety of novel authentication methods have been investigated. Traditional PIN-based mechanisms are notably vulnerable to shoulder-surfing attacks \cite{Tari2006Shoulder}. This issue is exacerbated in smart glasses, where the PIN is entered on a touchpad, making it more visible to bystanders \cite{Rasel2018GlassPass, GlassOTP2015}.
Alternative approaches such as voice-based PIN entry have been explored \cite{bailey2014typing, Yadav2015VBP, Li2021Gtalker}. These methods employ a cipher to map the PIN to random digits, obscuring the password from eavesdroppers. However, this technique necessitates mental computations from users, potentially diminishing usability \cite{bailey2014typing}. Camera-based authentication methods have also been a focus of research. The Glass OTP \cite{GlassOTP2015} system utilized the camera in Google Glasses to scan a QR code on the user's smartphone for authentication, although this method required the user to carry a smartphone. C-Auth \cite{lim2023cauth} used a downward-facing camera in the glasses to capture facial contours for authentication. Zhang et al. \cite{zhang2018continuous} developed a continuous authentication system based on eye movement response to visual stimuli, detected by a camera. Another approach involved iris recognition, where internal infrared cameras were used for authentication \cite{Yunghui2017Iris}. Behavioral biometrics have also been explored for smart glass authentication. Jagmohan et al. \cite{Jagmohan2016Gesture} demonstrated a gesture-based continuous authentication system. The GlassGuard system \cite{Ge2017GlassGuard}, utilized behavioral biometrics derived from touch gestures and voice commands for continuous authentication. Kawasaki et al. \cite{Kawasaki2022Blink} introduced an authentication method by observing the skin deformation during blinking, employing a photoreflector to measure blinks. Isobe et al. \cite{isobe2023personal} introduced an innovative approach using active acoustic sensing. This method involved transmitting acoustic signals through the nose using speakers integrated into the nose pads of the glasses, with the received signals captured by microphones. The acoustic structure of the nose served as a biometric identifier, though this technique could be affected by the dryness of the nose. Furthermore, this prototype was not tested across different days so the stability of the system was unclear.

\begin{table}[t]
\caption{SonicID and Other Authentication Methods on Wearables. Some papers reported Equal Error Rate (EER) which was converted to Balanced Accuracy ($BAC=1-EER$) in this table. The balanced accuracy in this table is the cross-session performance if the system was evaluated across different remounting sessions.}
\label{Tab: comparision with previous}
\footnotesize
\begin{tabular}{|c|c|c|c|c|c|c|c|c|}
\hline
\multirow{2}*{Project} & Form & \multirow{2}*{Sensors} & Biometrics & User & Cross- & Cross- & Study & Balanced\\ 
~ & Factor & ~ & Extracted & Interaction? & Session? & Day? & Participants & Accuracy\\
\hline\hline
\textbf{SonicID} & \textbf{Glasses} & \textbf{Acoustics} & \textbf{Face} & \textbf{No} & \textbf{Yes} & \textbf{Yes} &\textbf{40} & \textbf{96.6\%} \\ 
\hline
Isobe et al. \cite{isobe2023personal} & Glasses & Acoustics & Nose & No & Yes & No & 11 & 91.0\%\\ 
\hline
C-Auth \cite{lim2023cauth} & Glasses & Camera & Face & No & Yes & Yes & 20 & 96.5\%\\ 
\hline
Zhang et al. \cite{zhang2018continuous} & Glasses & Camera & Eyes & Yes & Yes & Yes & 30 & 93.1\%\\ 
\hline
Li et al. \cite{Yunghui2017Iris} & Glasses & Camera & Iris & No & No & No & 62 & 95.4\%-98.5\%\\ 
\hline
Kawasaki et al. \cite{Kawasaki2022Blink} & Glasses & Optical & Blink & Yes & No & No & 7 & 93.6\% \\ 
\hline
EarEcho \cite{gao2019earecho} & Earables & Acoustics & Ear Canal & No & Yes & Yes & 20 & 94.5\%\\ 
\hline
ToothSonic \cite{wang2022toothsonic} & Earables & Acoustics & Toothprint & Yes & No & No & 25 & 92.9\%-98.9\% \\ 
\hline
F\textsuperscript{2}Key~\cite{duan2024f2key} & Headphone & Acoustics & Mouth & Yes & Yes & No & 26 & 95.3\% \\ 
\hline
WristAcoustic \cite{Huh2023wristacoustic} & Wristband & Acoustics & Wrist & Yes & Yes & Yes & 25 & 95.0\% \\ 
\hline
\end{tabular}
\end{table}

\subsubsection{Other Wearable Devices}
In the domain of smartwatches, researchers have extensively explored biometric-based authentication methods. For instance, Cornelius et al. \cite{Cornelius2014BioImp} investigated the use of on-wrist bioimpedance responses for user authentication. Zhao et al. \cite{Zhao2020PPG} employed photoplethysmography signals for continuous authentication. Another innovative approach by Watanabe et al. \cite{Watanabe2021Acoustic} utilized active acoustic sensing, requiring users to perform four distinct hand poses for authentication. Further, Lee et al. \cite{Lee2021Vibration} proposed a method leveraging vibration responses, measured through accelerometer and gyroscope sensors, to authenticate users.
Recent advancements have introduced more sophisticated techniques. WristAcoustic \cite{Huh2023wristacoustic} constructed a classifier based on three hand poses, utilizing a cue signal emitted from a surface transducer and recorded by a contact microphone. Additionally, WristConduct \cite{sehrt2022wrist} innovatively used sound waves transmitted through the wrist bones, employing a bone conduction speaker and a laryngophone for authentication purposes.

In addition to on-device authentication methods, researchers have explored approaches that incorporate secondary devices to facilitate user authentication. Nymi band \cite{Nymi} is a wrist-worn wearable device that integrates fingerprint recognition, serving as an authentication mechanism for various devices and applications. Furthermore, ear-based authentication systems such as EarAE \cite{EarAE2023} and EarEcho \cite{gao2019earecho} utilized acoustic sensing techniques to capture the unique structure of the ear canal, thereby offering a biometric signature for authentication purposes. ToothSonic \cite{wang2022toothsonic}, employed an earable-based system that utilized toothprint-induced sonic waves for user verification. Similarly, Amesaka et al. \cite{amesaka2023user} have demonstrated the potential of using sound leakage signals from hearables to capture the acoustic characteristics of the ear canal, auricle, or hand, which can then be employed for authentication. EarDynamic \cite{wang2021eardynamic} leveraged acoustic sensing in earables to assess both the static and dynamic motions of the ear canal during speech, providing a novel method for authentication. F\textsuperscript{2}Key~\cite{duan2024f2key} extracted acoustic features associated with the mouth movements when users spoke on a headphone and used the distinctive mapping between the acoustic features and the utterance for authentication.

\subsection{Comparison between SonicID and Other Authentication Methods on Wearables}
\label{Sec: comparison with prior work}

To better compare our work with prior work of user authentication on wearables, we list all the projects that are most related to SonicID in Tab.~\ref{Tab: comparision with previous}. As shown in the table, SonicID does not need users' interaction to perform authentication and achieves comparable balanced accuracy, if not better, to prior work on glasses. This performance is evaluated with a valid number of participants and remains relatively consistent across different remounting sessions and days. In the meantime, SonicID maintains low-power and minimally-obtrusive due to the advantages of acoustic sensors. Specifically compared with existing vision-based authentication methods on glasses~\cite{lim2023cauth,zhang2018continuous,Yunghui2017Iris}, SonicID achieves better performance across different remounting sessions. Moreover, we believe that SonicID surpasses vision-based methods in terms of the sensor size, the power consumption and the resilience to environmental lighting conditions because acoustic sensors are known to be smaller, consume less power and be less sensitive to lighting conditions compared with cameras.

In addition to cameras, other sensors, e.g., photoreflectors and acoustic sensors, have been utilized to perform user authentication on glasses as well as earables, headphones and wristbands in prior work. They all achieve promising balanced accuracies but many of them require user interaction, such as blinking~\cite{Kawasaki2022Blink}, toothprint~\cite{wang2022toothsonic}, mouth movements~\cite{duan2024f2key} and hand poses~\cite{Huh2023wristacoustic}, for authentication, while SonicID automatically identifies users when the device is put on. EarEcho~\cite{gao2019earecho} also automatically scans users' ear canal with acoustic signals for authentication. However, the system is deployed on a pair of earphones and it captures the structure of the ear canal as biometrics, which are different from SonicID that captures the shape of the face as biometrics on a pair of glasses. 

Isobe et al. \cite{isobe2023personal} innovatively integrated acoustic sensors into the nose pads of a pair of glasses for user authentication. Though it reaches a promising performance within the same day, the system is easily affected by the nasal conditions because the acoustic sensors are in direct contact with users' nose. The experiments in the paper indicated that the balanced accuracy of the system drops from 91.0\% to 88.4\% and 83.0\% respectively if the participant's nose is wet or stuffed with cotton. Instead, SonicID places acoustic sensors on the hinges of the glasses, where many commercial smart glasses place their speakers and microphones. This helps the system suffer less from the impact of the internal state of users' body, which guarantees better stability in practice. In addition, the proposed system in \cite{isobe2023personal} operates at a sampling rate up to 96 kHz, which is not supported by every commodity device. By comparison, SonicID requires a sampling rate of 50 kHz. The narrower operating frequency band limits the amount of useful information our system is able to obtain but SonicID still achieves a better authentication performance while proposing a lower technical requirement. Compared with the proposed system in \cite{isobe2023personal}, it is easier for SonicID to be integrated into existing commodity smart glasses in the future because of its sensor position and technical requirements. In summary, SonicID contributes to the field by presenting the first acoustic-based method on smart glasses that scans users' faces to obtain biometric information for user identification.

%% file: 4_system_design.tex
\section{Facial Biometirc Information Extraction with Active Acoustic Sensing}
\label{Sec: Design}
In this section, we first introduce background on using face-based biometric information for authentication, followed by the core sensing technique and algorithm of SonicID to obtain unique acoustic features of each user. Then we present the deep learning model used in SonicID to authenticate users and the metrics used to evaluate the performance of our system.

\subsection{Face-based Biometric Authentication}
Biometrics are unqiue physical characteristics that can be used for personal recognition~\cite{biometrics}. Face, as one of the most important parts of human body, has been a crucial source from which biometric information is extracted because the geometries of the facial surfaces of different people are different. Extensive research has been carried out to utilize this biometric information from the user's face for authentication, especially captured by frontal cameras~\cite{beumier2000automatic,bicego2006on}. One of the most commonly used authentication methods based on facial biometrics captured by cameras is Apple's FaceID~\cite{faceid}. It delivers impressive performance in authentication and has advantages of being more convenient and secure compared with password-based methods. The core sensing pricinple of SonicID is similar to FaceID. It places acoustic sensors on smart glasses to scan the user's face in 3D with ultrasonic sound waves instead of infrared cameras and uses the obtained biometric information for authentication. 

To realize this goal, we decided to employ the calculation of echo profiles as the main data processing method because it exhibits promising performance in extracting patterns related to facial movements~\cite{li2022eario,li2024eyeecho} and body movements~\cite{mahmud2023posesonic} in prior work. Nevertheless, all these application scenarios are motion-related while our work aims to exploit echo profiles to represent the acoustic patterns of facial scanning and validate the feasibility of user identification based on these patterns. The main contribution of this paper is not to propose the echo profile calculation. Instead, we use echo profiles as a standard acoustic data processing technique, similar to mel-spectrograms and Doppler effects, and our goal is to explore how the acoustic patterns can be better exploited to serve the purpose of user identification. To reach this end, we designed several specific steps in the pipeline such as the selection of static instances (Sec.~\ref{Sec: echo profile calculation}) and the two-stage training technique for the machine learning model (Sec.~\ref{Subsec: ML}), to boost the authentication performance. Moreover, we experimented on the effectiveness of each step of the system pipeline including the choice of two pairs of speakers and microphones, the length and number of the static instances, and the pre-training stage in Sec.~\ref{Sec: Evaluation}. The evaluation results validated the effectiveness of the data processing and machine learning pipeline of SonicID. The major contribution of SonicID is to propose the first acoustic-based method on smart glasses that is able to perform user identification by scanning users' full face. The subsequent subsections introduce the principle of operation of SonicID in detail.

\subsection{Signal Transmission and Reception}
\label{Sec: fmcw}
SonicID adopts active acoustic sensing to scan a user's face as soon as they wear the device. Active acoustic sensing includes an emitter (e.g. speakers) to transmit encoded signals and a receiver (e.g. microphones) to receive the reflected signals. By comparing the received signals and the transmitted signals, one can obtain and analyze the status and change of the objects in the environment that the signals are targeted at. 

\begin{figure}[t]
    \centering
    \subfloat[Transmitted Signal (18-21 kHz)]{
        \includegraphics[height=.17\textwidth]{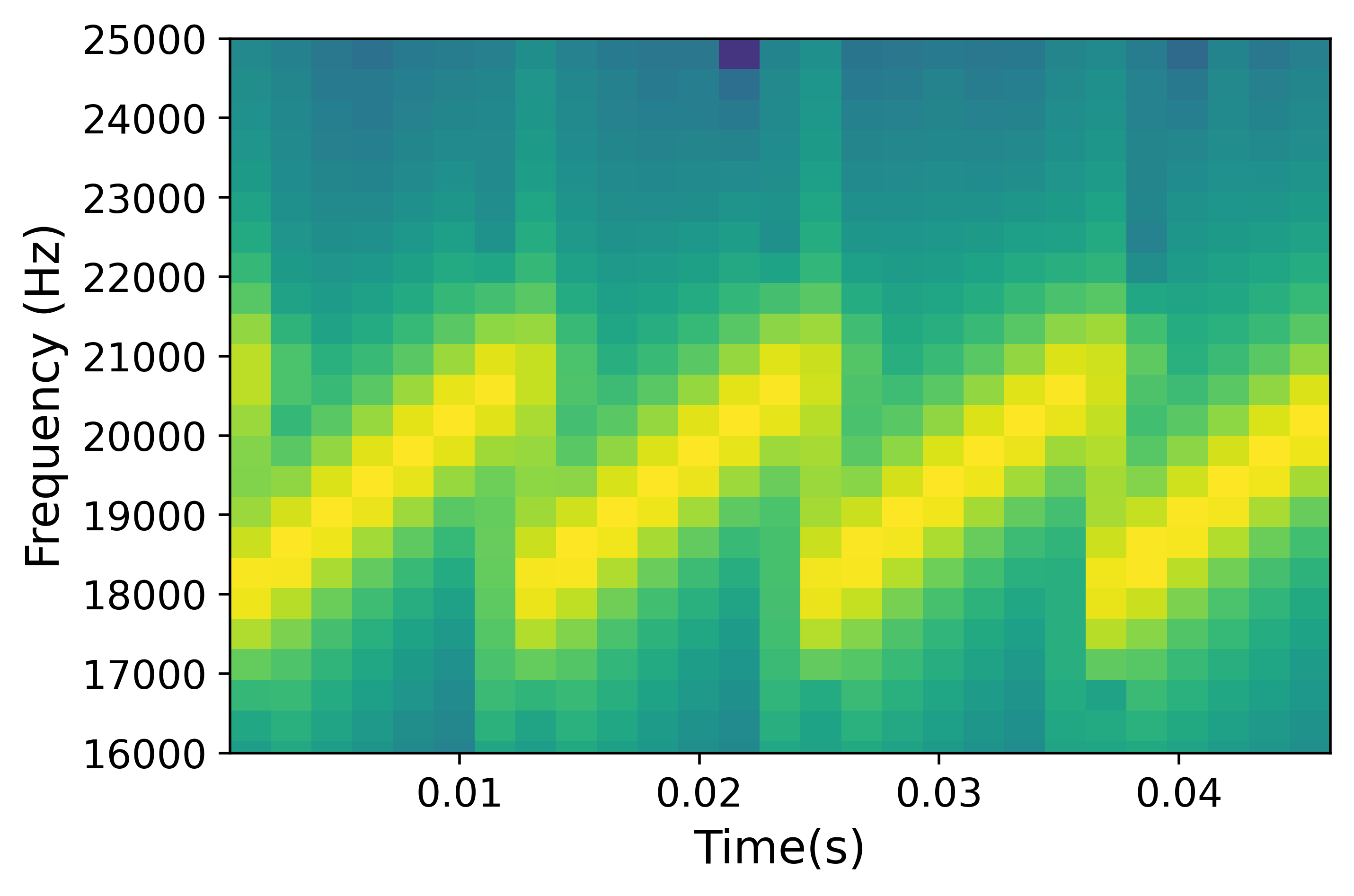}
    }
    \subfloat[Received Signal (18-21 kHz)]{
        \includegraphics[height=.17\textwidth]{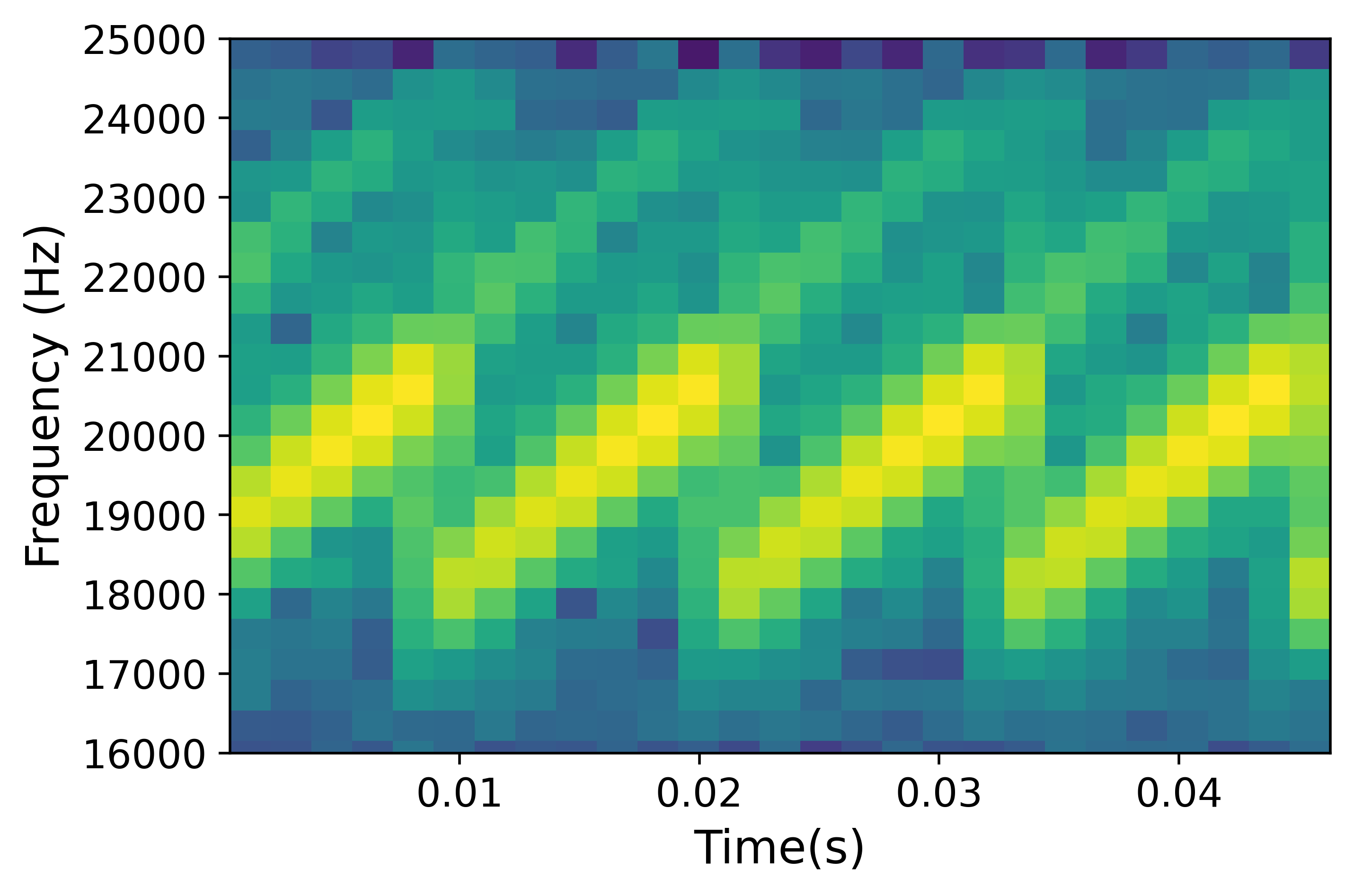}
    }
    \subfloat[Cross-correlation Result]{
        \includegraphics[height=.17\textwidth]{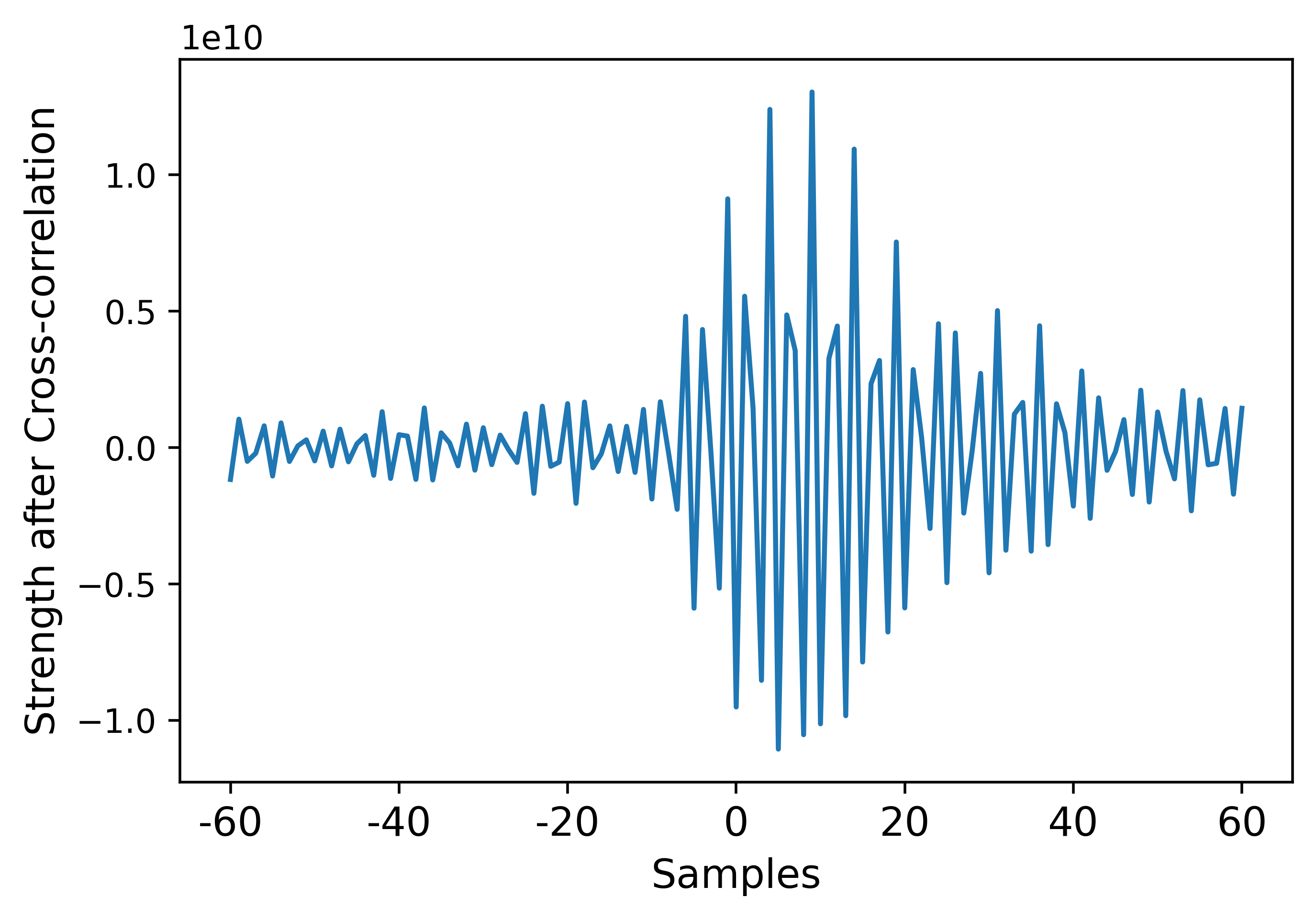}
    }
    \subfloat[Echo Profile]{
        \includegraphics[height=.17\textwidth]{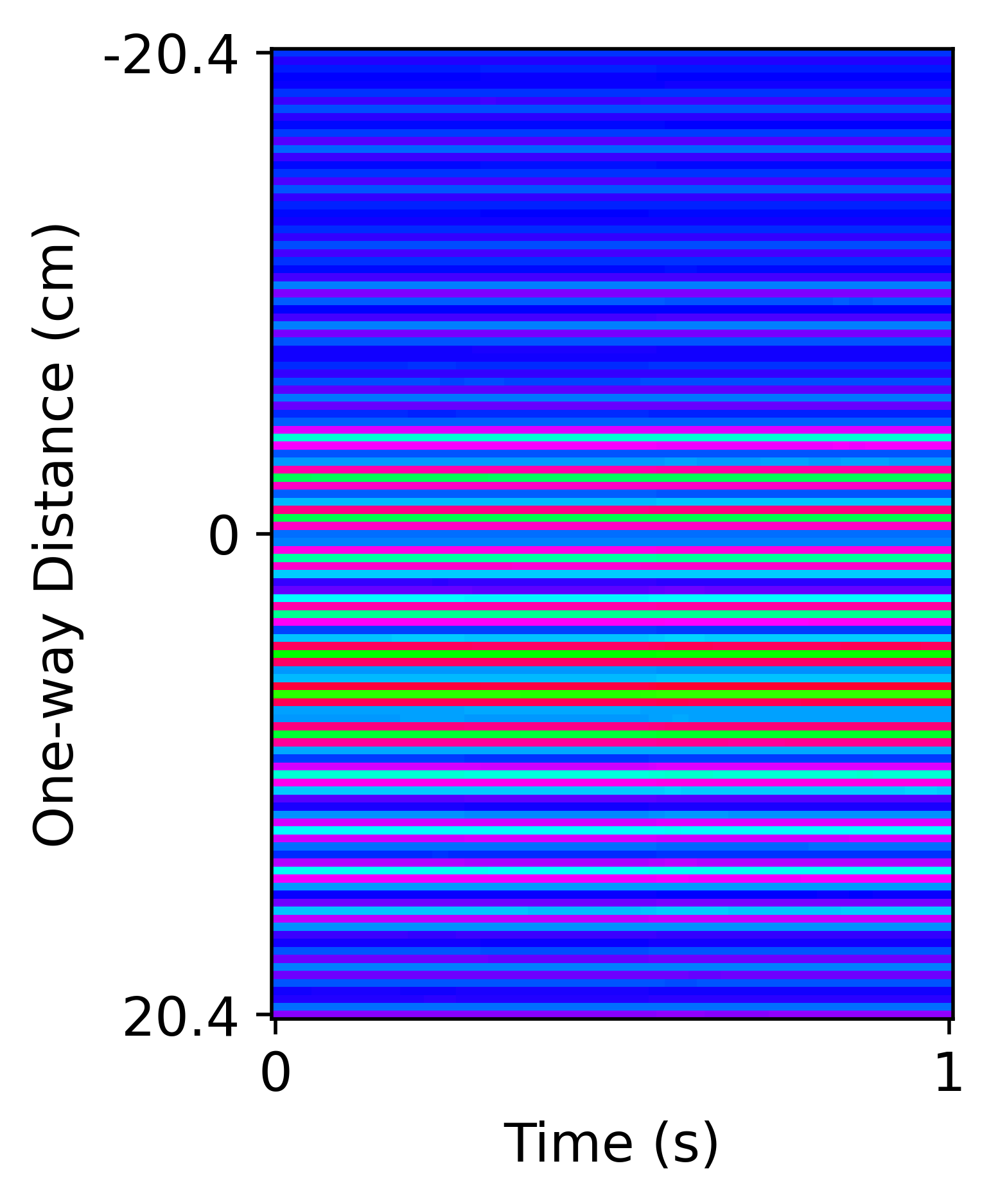}
    }
    \caption{FMCW Signal Transmission, Reception and Processing.}
    \label{Fig: signals}
\end{figure}

In SonicID, we choose signals that sweep within a certain frequency range as transmitted signals. Considering SonicID is deployed on smart glasses which is quite close to users when they put it on, we decide to sweep the signals in the ultrasonic frequency bands, i.e. over 18 kHz, to alleviate the impact of the audible sound on users' daily activities. In order to obtain richer information from users' face, we put one pair of speaker and microphone on each side of the glasses. The speakers on two sides are designed to transmit encoded signals in different ultrasonic frequency ranges to avoid interference. We pick 18-21 kHz for the speaker on the right side and 21.5-24.5 kHz for the speaker on the left so that they are inaudible to most people and compatible with the audio interfaces on most commodity devices. According to the Nyquist sampling theorem, the sampling rate of the system should be at least twice the highest frequency it supports to avoid alias. Hence, the lower bound of the sampling rate for our system is $2 \times 24.5 kHz = 49 kHz$. In practice, the sampling rate needs to be slightly larger than this lower bound to allow some room for potential frequency discrepancy. Therefore, a sampling rate of 50 kHz is selected to support these two frequency ranges. We believe that this sampling rate can be supported by the audio interfaces on most current commodity devices. A higher sampling rate such as 96 kHz or 192 kHz is not selected because we would like to keep the technical requirements of SonicID as low as possible to make it easier to be integrated into future smart glasses, as discussed in Sec.~\ref{Sec: comparison with prior work}. In our design, the signals sweep from the lower frequency boundary (18/21.5 kHz) to the higher frequency boundary (21/24.5 kHz) every 12 ms. Therefore, each sweep contains N = 600 samples ($50 kHz \times 12 ms$). Fig.~\ref{Fig: signals} (a) demonstrates one of the transmitted signals that sweeps from 18-21 kHz. These signals are commonly used and named as Frequency Modulated Continuous Waves (FMCW) in the field. 

In practice, the two speakers emit the encoded signals that sweep in two frequency ranges repeatedly towards the user's face when the user put on the device. The signals are reflected by the user's face and captured by the two microphones that are placed next to the speakers. By analyzing the differences between the received signals and transmitted signals, we can obtain unique acoustic features related to this specific user, which will thus be used for user identification. A detailed description of how SonicID generates the acoustic features is presented below.

\subsection{Echo Profile Calculation}
\label{Sec: echo profile calculation}
 The two microphones capture two channels of audio on two sides of the glasses. An example of the received signal in one channel (18-21 kHz) is shown in Fig.~\ref{Fig: signals} (b). After receiving the signals, we first apply two Butterworth band-pass filters, which are from 18-21 kHz and 21.5-24.5 kHz, on the signals separately to filter out the noises that are outside the frequency range of our interest. By applying two band-pass filters on two channels of signals, we get four channels of filtered signals, two of which represent the signals that travel on the same side of the face while the other two represent the signals that travel across the face from one side to the other side of the face. The acoustic patterns associated to the user's face that we would like to obtain are called Echo Profiles, which have been demonstrated in several previous papers~\cite{wang2018cfmcw,li2022eario,mahmud2023posesonic,li2024eyeecho}. According to prior work, we continuously calculate the cross-correlation between the received signals and the transmitted signals using Eq.~\ref{Eq: cross correlation}. 

\begin{equation}
\label{Eq: cross correlation}
    C(n) = Tx(n) * Rx(n) = \sum_{m=0}^{N-1} Tx(m) \cdot Rx(m+n), n \geq 0
\end{equation}
where $Tx(n)$ is the transmitted signal and $Rx(n)$ is the filtered received signal. Since there are four channels of filtered received signals after the band-pass filters, we calculate the cross-correlation between each one of these channels and the corresponding transmitted signal, leading to four channels of acoustic patterns related to different speaker-microphone links. Fig.~\ref{Fig: signals} (c) demonstrates $C(n)$ obtained after the cross-correlation calculation in one channel. The strength of $C(n)$ correlates to the strength of the signals that the system receives at different time. The 0-point is the direct path which means that the signal travels from the speaker directly to the microphone via solid since they are on the same glasses. The positive samples represent the signals that arrive at the system later than the direct path. They usually travel in the air and are reflected by the objects in the surroundings of the system. The negative samples represent those arriving earlier than the direct path and they come from previous frame(s) of the transmitted signals, which are being emitted continuously and repeatedly. Two consecutive samples are 0.02 ms apart ($1 \div 50 kHz$) and we only show the center 120 samples in Fig.~\ref{Fig: signals} (c).    

We then reshape the one-dimensional $C(n)$ by the frame size N = 600 samples to form a two-dimensional array. This array of patterns is defined as Echo Profile, which includes the biometric information of the face of the user who is wearing the device. Fig.~\ref{Fig: signals} (d) shows an example of the echo profile that is properly reshaped from Fig.~\ref{Fig: signals} (c). In echo profiles, the y-axis is the distance, which shows the strength of the signals that is reflected from different distances away from the SonicID system on glasses. Because different users have different facial shape and contours, the echo profile shows the facial scanning result of each user and is distinctive for each user. Considering the average length of people's faces, we crop 70 samples of echo profiles from -15 samples to 55 samples as the acoustic pattern to authenticate users instead of using the full range of it, which scans the area at a distance of 0 cm to 18.7 cm from the system ($55~samples \div 50 kHz \times 340 m/s \div 2$). We divide the distance by 2 because the signals travel round-way from the system to the face and back to the system. 18.7 cm is the maximum one-way distance from the system that we would like to scan. As stated above, the negative samples are from previous frames and theoretically do not contain much useful information. However, we still include 15 samples from the negative side because our transmitted signals are not infinite in the frequency domain and thus the acoustic patterns diffuse to the negative side to some extent when the cross-correlation is calculated. 

To achieve an optimal authentication performance, the system should obtain the scanned facial features of users when they are static to eliminate the impact of environmental factors. However, it is possible for the users to move and/or blink intentionally and/or unconsciously during the usage of the system, which might have a negative impact on the scanning results. Therefore, we decided to adopt multiple static short instances during which the users are mostly static to authenticate them instead of using one long instance. Based on our pilot study, we empirically chose five of the aforementioned 600-sample frames as one instance, which corresponds to 0.06 seconds in time ($5 \times 600~samples \div 50kHz$). In our user study, we will divide one session of data into multiple 0.06-second instances to evaluate the system. These instances of cropped echo profiles are fed into a deep learning model to authenticate users, which is introduced in the subsection below.

\subsection{Machine Learning Model for Authentication}
\label{Subsec: ML}
\begin{figure*}[t]
  \includegraphics[width=0.9 \textwidth]{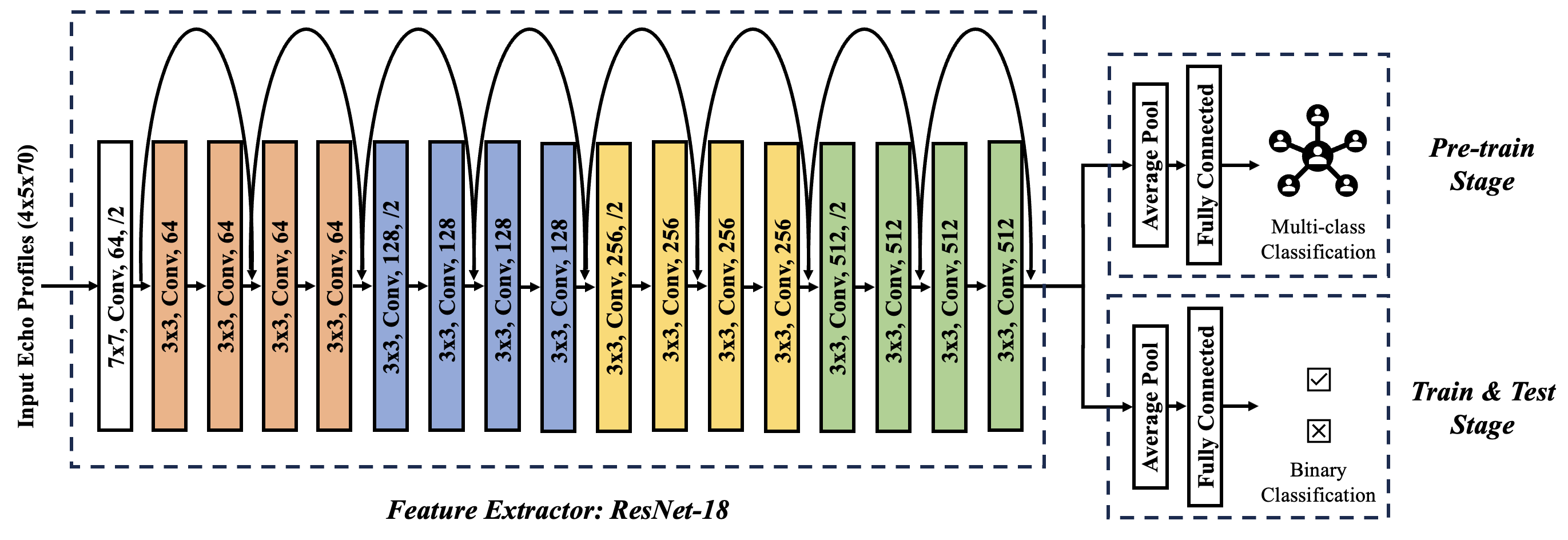}
  \caption{Machine Learning Pipeline for SonicID.}
  \label{Fig: ml}
\end{figure*}

Though the acoustic data is originally one-dimensional, we reshape the echo profiles to make them two-dimensional images. In order to extract features that are unique to each user from these two-dimensional echo profiles, we decide to adopt a deep learning model based on the ResNet-18 architecture because Convolutional Neural Networks (CNN) are known for being good at distinguishing different classes based on visual representations. Besides, ResNet-18 has a proper amount of parameters (around 11 million parameters), which is suitable for our dataset size and potential embedded on-chip deployment. Hence, we construct the model using the ResNet-18 architecture as a feature extractor to extract features from echo profiles and using a fully-connected layer as a classifier to distinguish samples from different classes, as shown in Fig.~\ref{Fig: ml}. We adopt a two-stage training method to optimize the system performance. First, we created a dataset from 21 users before the formal user study. This dataset is used to pretrain a base model, with the feature extractor and the classifier combined, to tackle a task of multi-class classification. By aiming to classify these 21 users with the highest accuracy, a powerful feature extractor is trained to extract distinguishable features from the acoustic patterns obtained by scanning each user's face. The base model only needs to be trained once and can be used for the individual models of all new users. In the second stage, the trained feature extractor is retained while the multi-class classifier is replaced with a binary classifier. The model, after further training, classifies the real user as positive and the attackers as negative. While training the individual model for a new user, the positive instances come from the data that this user provides when he/she registers on the device and the negative instances come from the existing dataset which was pre-collected from the 21 different users. This two-stage training technique enhances the feature extractor to focus more on the delicate differences across various users, compared with directly training a binary classifier.

As introduced in Sec.~\ref{Sec: echo profile calculation}, the input fed into the model is the reshaped and cropped echo profiles. We take 5 600-sample frames as one instance. Therefore, the size of each instance as input is $4 \times$ 5 $\times 70$, where 4 represents the 4 channels of audio data and 70 represents the number of the cropped vertical samples. During testing, the positive instances are still provided by the real user while the negative instances come from the unknown attackers that have not been seen by the model during training.

In the CNN model, we use an Adam optimizer, a cosine scheduler and set the initial learning rate to be 0.0002. The model is trained to minimize the cross-entropy loss for 20 epochs for the base model and for another 10 epochs for the individual models. Random vertical shift is included and random noise is added as the data augmentation methods while the model is trained to boost its generalizability. Both of these data augmentation methods are performed at the instance level. For each instance, we implement random vertical shift by first cropping 110 vertical samples and randomly selecting 70 consecutive samples out of these 110 samples to mitigate the impact of different wearing positions of the glasses. Theoretically, the augmented dataset is 40 times ($110 - 70$) larger than the original dataset after random vertical shifts if the model is trained for enough epochs. Random noise is introduced by multiplying each data point of the input instance with a random factor from 0.9 to 1.1 with a probability of 80\%. Moreover, z-score normalization is applied on each instance with a probability of 50\%. This method improves the generalizability of the model across different wearing sessions and different days while guaranteeing that there are enough original training samples in the meantime. With the existence of random noise, the augmented dataset is at least two times larger (normalized/non-normalized).

\subsection{Evaluation Metrics}
\label{Subsec: metrics}
To evaluate the performance of our model in SonicID, we adopt three commonly used metrics in prior authentication research, which are True Positive Rate (TPR), False Positive Rate (FPR) and Balanced Accuracy (BAC). The euqations to calculate these three values are as follows.

\begin{equation}
\label{Eq: tpr}
    True~Positive~Rate~(TPR) = \frac{True~Positives}{True~Positives+False~Negatives}
\end{equation}

\begin{equation}
\label{Eq: fpr}
    False~Positive~Rate~(FPR) = \frac{False~Positives}{False~Positives+True~Negatives}  
\end{equation}

\begin{equation}
\label{Eq: ba}
    Balanced~Accuracy~(BAC) = \frac{TPR+(1-FPR)}{2}
\end{equation}

TPR evaluates the ability of the system to authenticate the real users while FPR evaluates the ability of the system to protect itself from being attacked. BAC gives a balanced evaluation between the two metrics above.

%% file: 5_form_factor.tex
\section{Prototype Design and Implementation}
\label{Sec: form factor}
This section presents the design and implementation of the hardware prototype. First, we introduce the micro-controller and sensors used in the system. Next, we present how we prototype the system on the glasses form factor.

\begin{figure}[t]
    \centering
    \subfloat[Hardware Components]{
        \includegraphics[height=.19\textwidth]{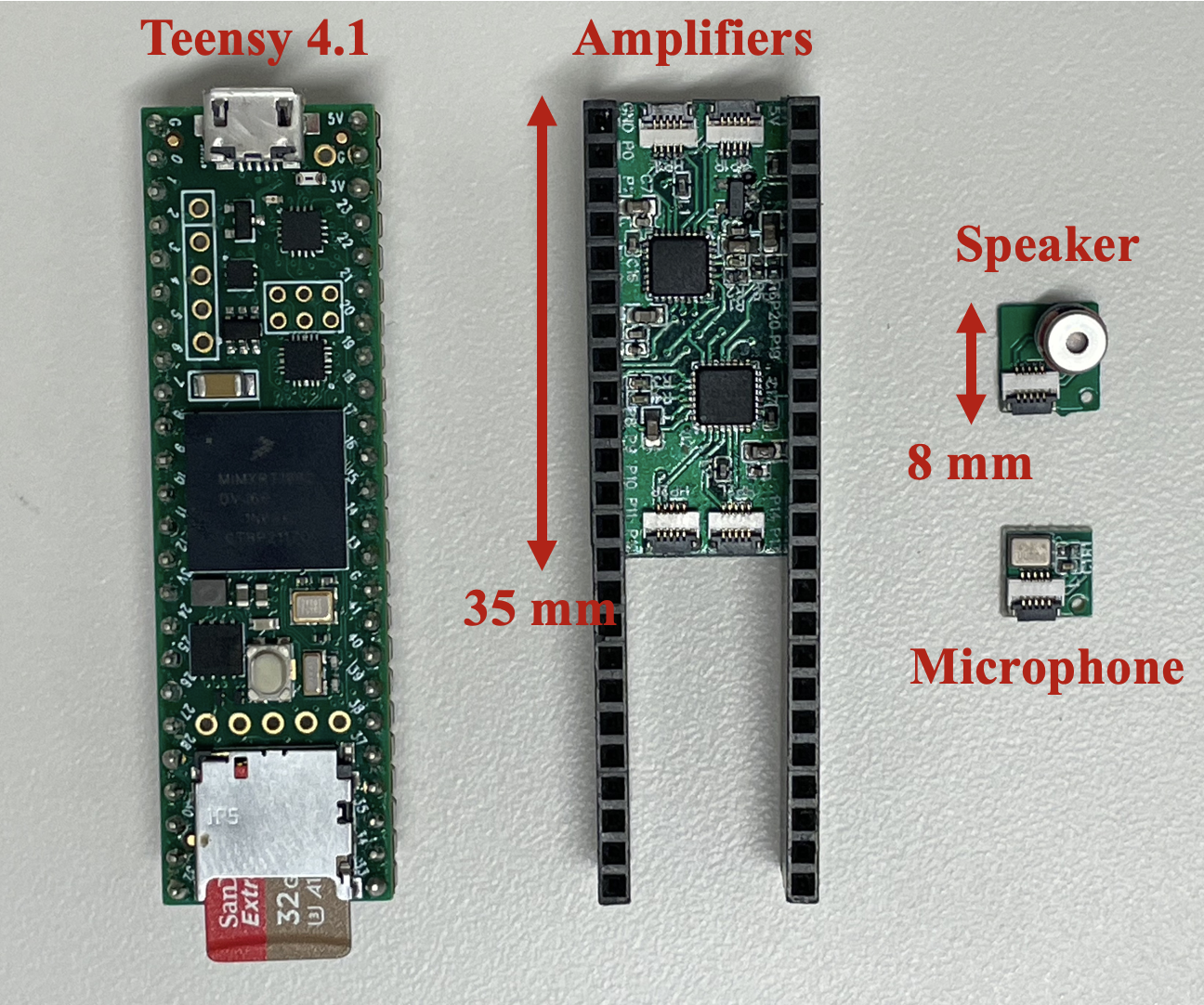}
    }
    \subfloat[Explore Sensor Positions]{
        \includegraphics[height=.19\textwidth]{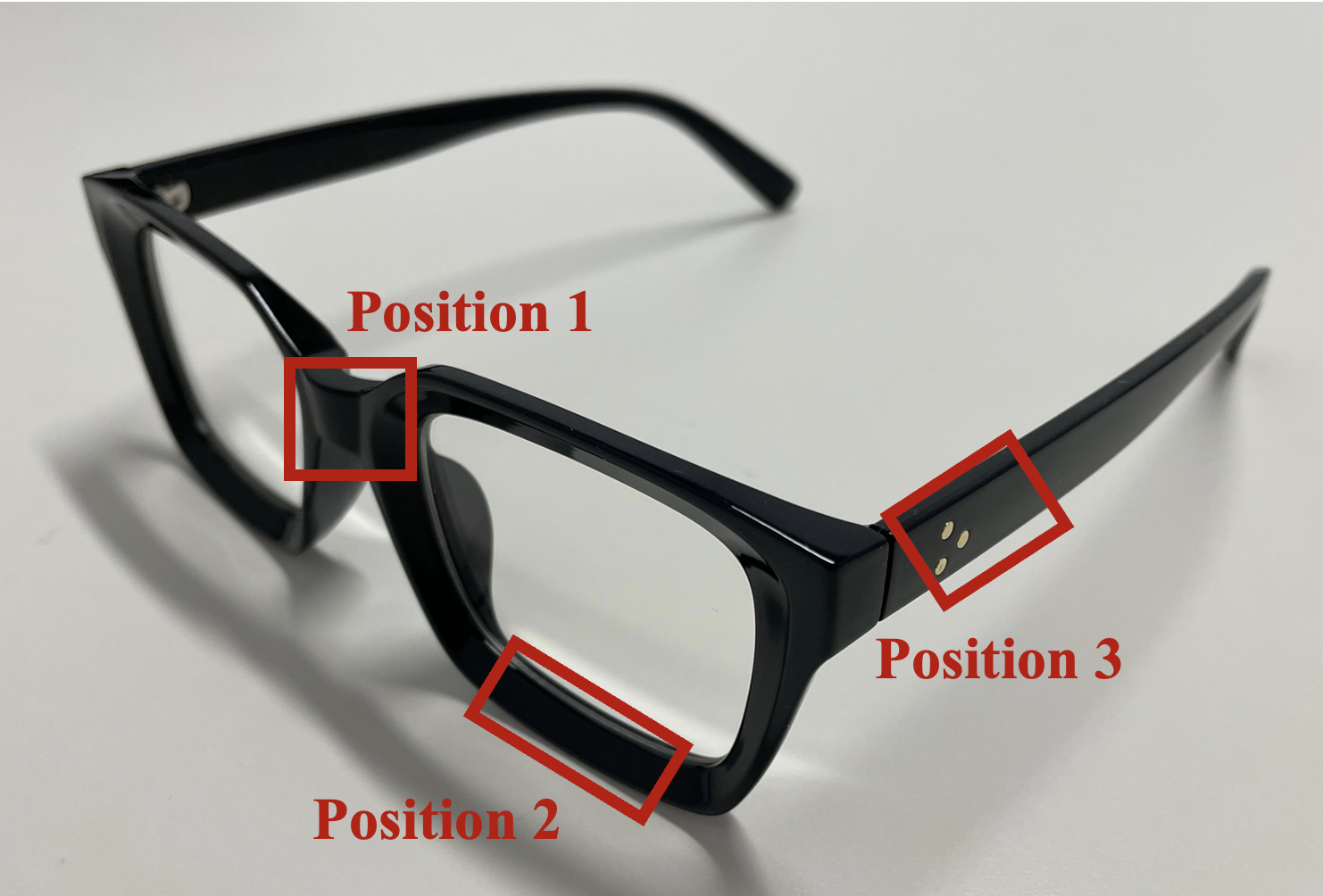}
    }
    \subfloat[Prototype]{
        \includegraphics[height=.19\textwidth]{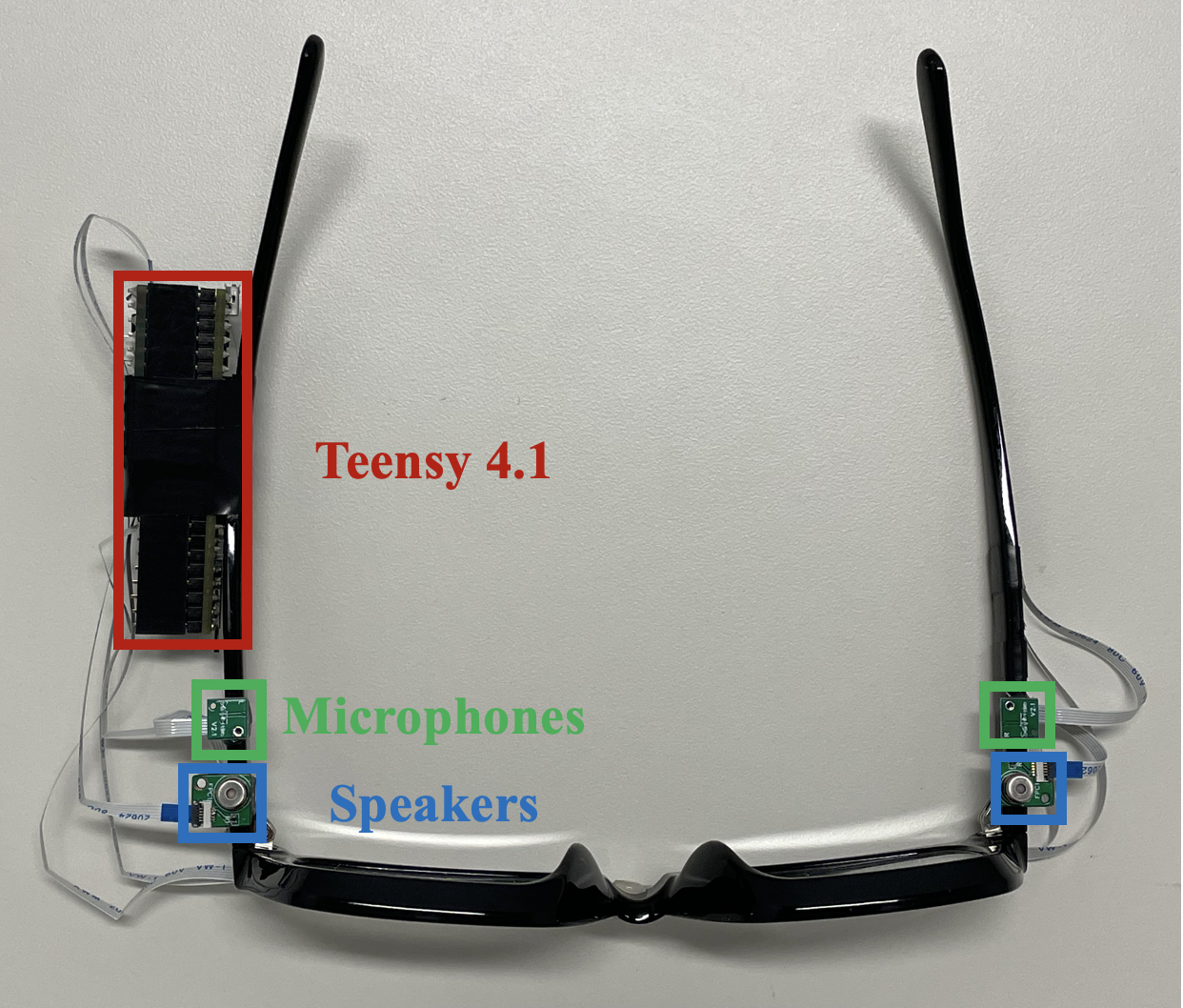}
    }
    \subfloat[User Wearing Prototype]{
        \includegraphics[height=.19\textwidth]{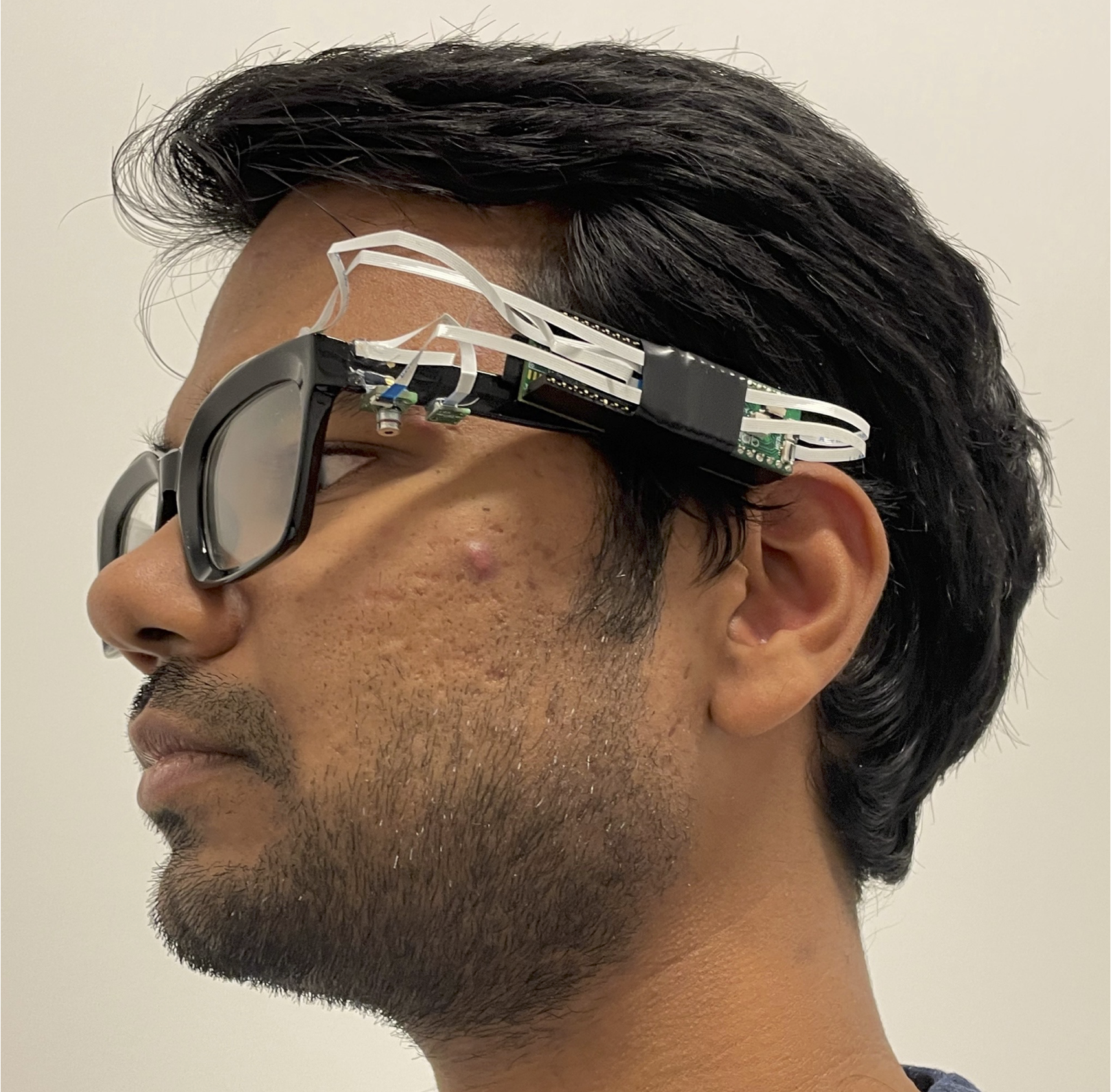}
    }
    \caption{Hardware Components and Prototype of SonicID.}
    \label{Fig: hardware}
\end{figure}

\subsection{MCU and Sensors}
Fig.~\ref{Fig: hardware} (a) demonstrates the hardware components used in the SonicID system. In order to implement the signal transmission and reception methods introduced in Sec.~\ref{Sec: fmcw}, we use the speaker OWR-05049T-38D\footnote{\url{https://www.bitfoic.com/detail/owr05049t38d-14578}} to emit encoded signals and the microphone ICS-43434\footnote{\url{https://invensense.tdk.com/products/ics-43434/}} to receive reflected signals. We customize Printed Circuit Boards (PCB) for the speakers and microphones to make them minimally-obtrusive and compatible with different micro-controllers. Teeny 4.1\footnote{\url{https://www.pjrc.com/store/teensy41.html}} is utilized as the micro-controller in the SonicID system to manage the operation of speakers and microphones because of its good performance in processing audio data. Two of the amplifiers MAX98357A\footnote{\url{https://www.analog.com/en/products/max98357a.html}} are used in the system to support as many as 2 speakers and 8 microphones. Specific PCBs are also designed for these amplifiers to make them easily fit the Teensy 4.1 board. Speakers and microphones are connected to Teensy 4.1 via Flexible Printed Circuit (FPC) cables and they communicate with each other with the Inter-IC Sound (I2S) buses. The transmitted signal is pre-programmed into the memory of Teensy 4.1 and the received audio data is stored into the on-board SD card on Teensy 4.1. 

\subsection{Form Factor}
After we select the appropriate components for the SonicID system, the next step is to prototype SonicID on the glasses form factor. In order to validate the feasibility of SonicID, we decide to deploy the system on a pair of commodity glasses. Since we want to scan the user's face to obtain their biometric information, we would like to point the speakers and microphones downwards, emitting signals directly towards the user's face. There are only limited space on glasses where we can place the sensors, facing downwards, as shown in Fig.~\ref{Fig: hardware} (b): (1) the bridge of the glass frame; (2) the bottom of the glass frame; (3) the hinges of the glasses. If the sensors are placed at position (1) or (2), there is a chance that the sensors may touch the user's face or block the view of the user because these two positions are right in front of the user's face. Therefore, we determine to put the sensors at the hinges of the glasses, where there is some space to instrument sensors without letting them touch the user's face or block the user's view. Moreover, this sensor position has been leveraged in multiple prior work to place acoustic sensors to track facial expressions \cite{li2024eyeecho} and body poses \cite{mahmud2023posesonic}. We can potentially integrate SonicID into their systems to realize activity tracking and user authentication with one set of hardware.

As displayed in Fig.~\ref{Fig: hardware} (c), we put one pair of speaker and microphone on each side of the glasses to obtain richer information and the speaker and microphone are close to each other to better capture the reflected signals. The micro-controller Teensy 4.1 is also fixed to the left leg of the glasses with hot glue and tapes to make the entire system wearable. The FPC cables connecting Teensy 4.1 and the speakers and microphones are properly restrained and aligned to the glass frame. The final prototype looks similar to a pair of commercial smart glasses and is comfortable to wear, as shown in Fig.~\ref{Fig: hardware} (d). We used a scale to measure the weight of the prototype. The prototype itself is weighed to be 48.5 grams in total while the pair of commercial glasses we purchased are 27.8 grams and the MCU and sensors on it are 20.7 grams. If a Li-Po battery of 150 mAh capacity is included into the system, the total weight of the prototype will be 52.3 grams, which we assume is comfortable to wear for general users.

%% file: 6_user_study.tex
\section{User Study}
\label{Sec: user study}

To evaluate the performance of SonicID, we designed and conducted a user study across several days and various scenarios. The goal of the study is to validate SonicID's capability of authenticating users and protecting the system from attacks.

\subsection{Study Design}
We used the prototype described in Sec.~\ref{Sec: form factor} to conduct the user study. In addition to the prototype, a laptop (MacBook Pro, 13-inch, Apple M1 chip) was used in the study to show the instruction video to the participants. The study was conducted in an experiment room on campus. When the study started, the participants came in and was introduced the detailed study procedures before signing the consent form. Then they sat in front of the laptop and wore the prototype. Some participants had long hair which sometimes got stuck between the hardware components and the glasses when they wore the device so we provided hair ties for participants to put their hair up if their hair was long. 

After the prototype was properly worn, the data collection process started. An instruction video was played on the laptop. The participants were first instructed to sit still for 10 seconds when the system scanned their face continuously and then asked to remount the device within 10 seconds. Remounting meant taking off the prototype and putting it back on. This procedure was designed to simulate real-life wearing experience since users of wearable devices frequently remount the device everyday. This added variance to the collected data. After the prototype was remounted, the participants stayed still for another 10 seconds and repeated the process above. The data collection lasted for 12 minutes for each participant where 36 10-second remounting sessions of data was collected. For a selected number of participants, after the data collection process above, we asked them to repeat the aforementioned process for 2 minutes respectively when they were talking, standing and lying down. The participants first kept talking while the system scanned their face. Then they stood up and stood still while the system was in operation. Lastly, they lied down on a bean bag to evaluate the system. This design aimed to evaluate the performance of SonicID in various application scenarios.

All the participants were asked to come back on two other days after the first day and collected data two more times with the same procedures because we would like to evaluate the stability of SonicID across different days. After the participants were done with the study on all three days, they filled out a questionnaire to provide their demographic information and feedback on using the prototype and the system.

\subsection{Participants}
The user study was approved by the Institutional Review Board (IRB) of the home institution of the researchers. We successfully recruited 40 participants (21 females, 17 males and 2 non-binaries) with an average age of 22.6 years old, ranging from 18 years old to 30 years old. During the recruitment of participants, we did not set any criteria to exclude any specific participant except that they should be at least 18 years old. The set of participants is mostly gender balanced and covers diverse ethnic groups. Furthermore, SonicID utilizes a sound-based technique to authenticate users so the skin colors of participants should not make a large difference in terms of system performance in theory, compared with camera-based methods. As a result, we anticipate that SonicID is capable of being scaled to a much larger group of users with different backgrounds.

For each participant, we collected 18 minutes of data while they were sitting, which were divided in 108 10-second remounting sessions across three days. For the last 16 participants, we collected another 9 minutes of data from each participant when they were in other body postures, which included 54 10-second remounting sessions across three days. Hence, 864 minutes of data was collected in total for this user study. Note that the data P11 collected on Day 3 was lost due to the hardware issue so we asked the participant to come back again and redo the study on Day 3.

%% file: 7_results.tex
\section{Evaluation}
\label{Sec: Evaluation}
In this section, we present the evaluation results of the user study stated in Sec.~\ref{Sec: user study}. We compare the performance of SonicID with different settings. Moreover, we evaluate the usability of the system by analyzing the responses to the questionnaires collected after the study.

\subsection{Study Results}
\label{Sec: study results}
\begin{figure*}[t]
  \includegraphics[width=1 \textwidth]{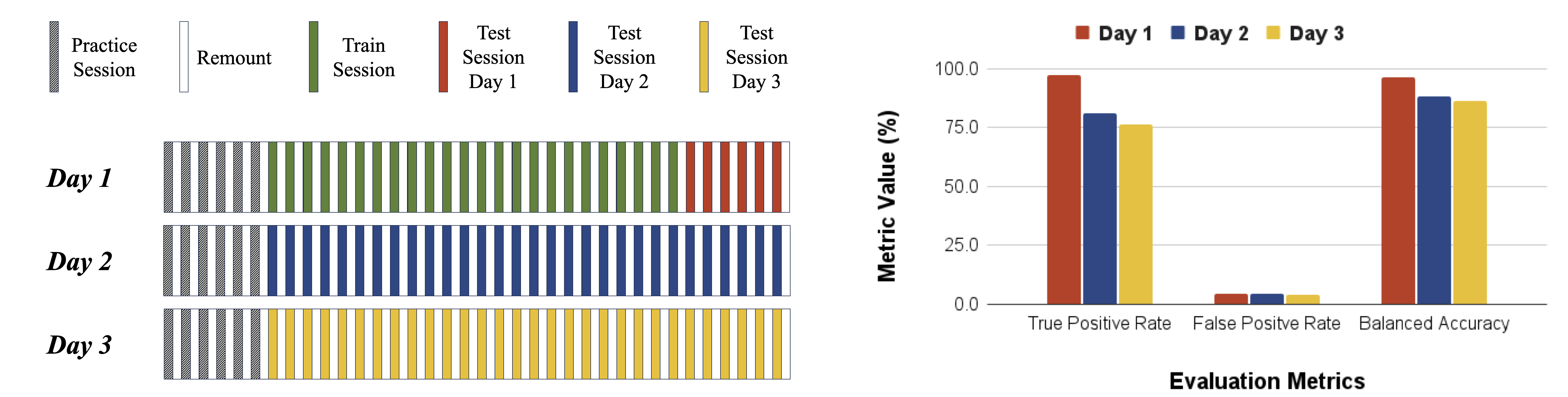}
  \caption{Evaluation Setup and Average Results of 40 Participants across Sessions and Days.}
  \label{Fig: setup and results}
\end{figure*}

In this subsection, we analyzed the performance of SonicID across multiple remounting sessions and days to simulate real-world wearing experience of smart glasses. The evaluation setup and results are visualized in Fig.~\ref{Fig: setup and results}.

\subsubsection{Cross-session Performance}
In the user study, we collected 36 sessions (6 minutes) of data from each participant while they were sitting still on each day. We first evaluated SonicID's performance across different remounting sessions on the same day. Out of the 36 sessions we collected on Day 1, we considered the first 6 sessions (1 minute) of data as the practice sessions and dumped them while we evaluated the system. Among the remaining 30 sessions, we used the first 24 sessions (4 minutes) of data as positive samples to train the deep learning model described in Sec.~\ref{Subsec: ML}. According to the data processing pipeline introduced in Sec.~\ref{Sec: echo profile calculation}, we divided each 10-second session into multiple instances of 0.06 s and empirically picked top ten instances with the smallest movements by comparing the energy change within these instances. Ten 0.06-second instances from each session were actually used for training the model. The negative samples used to train the model came from a dataset we collected prior to the user study. To collect this dataset, we ran a pilot study of the same procedures as the formal user study with 21 people, including 7 researchers and 14 people from their networks. Among these people, 9 of them collected data on three days and the remaining 12 collected data on one day. Each person contributed 5 or 6 minutes of data of sitting still on each day. All these data in the dataset were used for training the base model in the pretraining stage and the individual models as negative samples. The same process was conducted to select the most static instances from all the sessions while training.

After the model was properly trained for each participant in the user study, the remaining 6 sessions (1 minute) of data from this participant was adopted as positive samples to test the model. The negative samples for testing came from the other 39 participants in the user study. Each of these 39 participants contributed all sessions of data, excluding the first 6 sessions which were considered as practice sessions, to test the model. In this way, these "attackers" during evaluation were not seen by the model in the training process. Static instances were also selected from all sessions for testing. The model made binary classifications to authenticate users and computed the three metrics introduced in Sec.~\ref{Subsec: metrics} to evaluate the results. We trained an individual model for each participant and the average TPR, FPR and BAC across 40 participants is 97.2\% (std=9.0\%), 4.6\% (std=4.7\%) and 96.3\% (std=4.9\%). The detailed metrics for every participant are shown in Figs.~\ref{Fig: results tpr} to \ref{Fig: results ba} respectively in Appendix~\ref{Appendix: individual results}. 

The average FPR of 4.6\% indicates that the system rejects false users with a success rate of over 95\%. It demonstrates the ability of SonicID to protect the smart glasses from being attacked by bad actors. The average TPR of 97.2\% means that the system authenticates the real users successfully with a rate of over 95\%. It shows the ability of SonicID to recognize true users of the smart glasses. BAC gives a balanced evaluation of the two metrics above. 

\subsubsection{Cross-day Performance}
The results above validated the performance of SonicID across different remounting sessions on the same day. Then we evaluated the performance across different days, which was meant to validate the stability of SonicID. On Day 2 and Day 3, we also collected 6 minutes of data from each participant when they were sitting still using the same procedures as Day 1. To be consistent with Day 1, we also dropped the first 1 minute of data from every participant. Then the static instances picked from the remaining data were used for testing the cross-day performance of SonicID. We did not retrain the models and directly used the models trained on Day 1. Each participant provided his/her own 5 minutes of data as positive samples to test the model for TPR on Day 2 and Day 3 separately. The negative samples still came from the other 39 participants in the user study to test FPR.

The results showed that the average TPR, FPR and BAC for Day 2 across 40 participants are 81.3\% (std=27.8\%), 4.7\% (std=4.5\%) and 88.3\% (std=14.5\%) while those for Day 3 are 76.6\% (std=29.2\%), 4.0\% (std=3.6\%) and 86.3\% (std=15.2\%). The individual study results for each participant are displayed in Figs.~\ref{Fig: results tpr} to \ref{Fig: results ba} respectively in Appendix~\ref{Appendix: individual results}. We ran a one-way repeated measures ANOVA test among TPR on three days for all 40 participants and identified a statistically significant difference ($F(2, 117) = 11.39, p = 0.00005 < 0.05$). Specifically, we conducted a repeated measures \textit{t}-test between TPR on Day 1 and Day 2, and we found a statistically significant difference ($t(39)=3.50, p=0.0012<0.05$). While the same repeated measures \textit{t}-test was conducted between TPR on Day 1 and Day 3, a statistically significant difference was found ($t(39)=4.29, p=0.0001<0.05$). This indicates that TPR decreases on both Day 2 and Day 3, compared to that on Day 1. However, the average TPR is still over 75\% even on Day 3. Under the cases when the real users fail to authenticate themselves into the smart glasses, they can adjust the glasses position or remount the device once, and they will likely log into the system successfully. The decrease of TPR can be attributed to the lack of positive training data from the user. One potential solution to improving TPR is to continuously collect the biomentric information of users' face while they are using the device and update the model accordingly to increase the success rate of authentication. We explored this method in Sec.~\ref{Sec: fine tune}.

Similarly, we also ran a one-way repeated measures ANOVA test among FPR on three days for all 40 participants and did not find a statistically significant difference ($F(2, 117) = 2.70, p = 0.07 > 0.05$). This validates the stability of SonicID to protect the system from attackers across different days. In such authentication systems, it is usually more crucial to keep FPR lower than keeping TPR higher in case of a tradeoff since we generally do not want other people to have easy access to our personal devices.

In Fig.~\ref{Fig: results tpr} and Fig.~\ref{Fig: results fpr}, we can see that P32 has the lowest TPR on Day 2 and P30 has the lowest TPR on Day 3 while P4 has the highest FPR on Day 2 and P27 has the highest FPR on Day 3. By checking the videos of the participants recorded during the study for reference, we figured that P4, P27, P30 and P32 all have long hair and bangs which made it harder for them to put the glasses back to the same position every time after remounting. This negatively affected the performance of the system on them. In Sec.~\ref{Subsec: improve performance}, we discussed several potential methods to improve the system performance and stability in the future.

\subsection{Impact of Different Amounts of Training Data}
\label{Sec: training data amount}

\begin{figure}[t]
    \centering
    \subfloat[True Positive Rate]{
        \includegraphics[height=.22\textwidth]{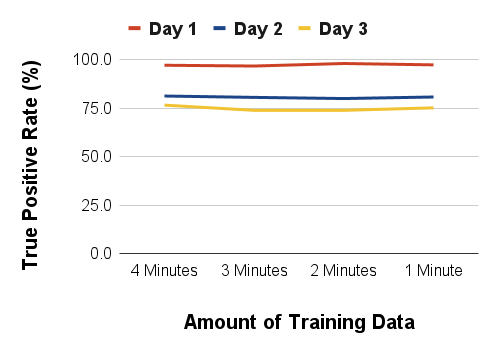}
    }
    \subfloat[False Positive Rate]{
        \includegraphics[height=.22\textwidth]{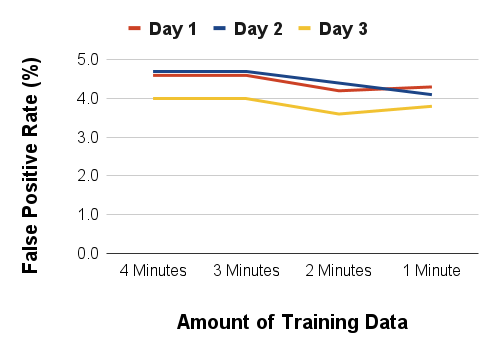}
    }
    \subfloat[Balanced Accuracy]{
        \includegraphics[height=.22\textwidth]{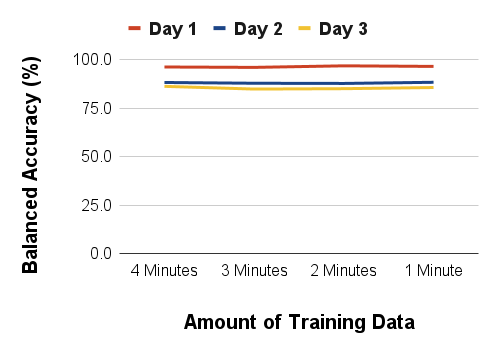}
    }
    \caption{Average TPR, FPR and BAC across 40 Participants with Different Amounts of Training Data.}
    \label{Fig: training data}
\end{figure}

In the evaluation results above, we adopted 4 minutes of training data from each participant, which leads to satisfactory performance. However, in practice, the fewer data a new user needs to provide to register on the device, the more user-friendly the system will be. Therefore, we conducted another experiment to explore the impact of different amounts of training data on the system performance. To evaluate this factor, we discarded the last 2.5 seconds, 5 seconds and 7.5 seconds of data from each 10-second session and still picked ten most static 0.06-second instances from the remaining data in each session to train the model. In practice, this reduces the amount of training data that each new user needs to provide from 4 minutes to 3 minutes, 2 minutes and 1 minute respectively. The average TPR, FPR and BAC among 40 participants with different amounts of training data is illustrated in Fig.~\ref{Fig: training data}. As shown in the figure, all of the three metrics remain similar across three days when the amount of training data decreases from 4 minutes to 1 minute. We ran three one-way repeated measures ANOVA tests among TPR with different amounts of training data for Day 1, Day 2 and Day 3 and did not find a statistically significant difference on any of these days ($F(3, 156) = 1.37, p = 0.26 > 0.05$ for Day 1, $F(3, 156) = 0.38, p = 0.77 > 0.05$ for Day 2, and $F(3, 156) = 1.82, p = 0.15 > 0.05$ for Day 3). The same ANOVA tests were performed among FPR and no statistically significant difference was found on any of the three days ($F(3, 156) = 1.32, p = 0.27 > 0.05$ for Day 1, $F(3, 156) = 2.35, p = 0.08 > 0.05$ for Day 2, and $F(3, 156) = 1.33, p = 0.27 > 0.05$ for Day 3). The statistical analysis validates that the system performance maintains satisfactory with the amount of training data as little as 1 minute from each new user. Therefore, we adopted 1 minutes of data from each participant on each day to train the model and evaluate the performance in all the experiments below.

\subsection{Leave-One-Day-Out Evaluation}
\label{Sec: leave on day out}
In the experiments above, we utilized participants' data on Day 1 to train the model and the data on Day 2 and Day 3 to test the model. In order to eliminate the impact of random factors among different days, we conducted leave-one-day-out experiments by using data from Day 2 and Day 3 to train the model respectively while the data from the other two days was used for evaluation. The experiment results are demonstrated in Tab.~\ref{Tab: different days}.

\begin{table}[b]
\caption{Average TPR, FPR and BAC across 40 Participants with Models Trained on 1-Minute Data from Different Days. Data Format: TPR/FPR/BAC. Results with Training and Testing Data from the Same Day are in Bold.}
\label{Tab: different days}
\begin{tabular}{| c | c | c | c |} 
\hline
 Training Data Source & Day 1 & Day 2 & Day 3\\ [0.5ex] 
 \hline\hline
 Train on Day 1 & \textbf{97.4\%/4.3\%/96.6\%} & 80.8\%/4.1\%/88.4\% & 75.2\%/3.8\%/85.7\%\\
 \hline
 Train on Day 2 & 80.0\%/3.7\%/88.1\% & \textbf{96.5\%/3.6\%/96.5\%} & 82.7\%/3.6\%/89.6\%\\
 \hline
 Train on Day 3 & 76.4\%/3.8\%/86.3\% & 82.7\%/3.2\%/89.7\% & \textbf{98.2\%/3.1\%/97.5\%}\\
 \hline
\end{tabular}
\end{table}

By averaging the leave-one-day-out evaluation results, the obtained average TPR, FPR and BAC when the system is evaluated on the same day as the training data is collected are 97.4\%, 3.7\% and 96.9\% while those metrics are 79.6\%, 3.7\% and 88.0\% on a different day. As a result, SonicID achieves a balanced accuracy over 95\% across different remounting sessions and that over 85\% across different days with a false positive rate under 4\%. These results again validated the capability and stability of SonicID especially in terms of protecting users' personal devices from attackers.

\subsection{Performance Boost across Days}
\label{Sec: fine tune}
\begin{figure}[t]
    \centering
    \subfloat[Reflected Signals from P25]{
        \includegraphics[height=.45\textwidth]{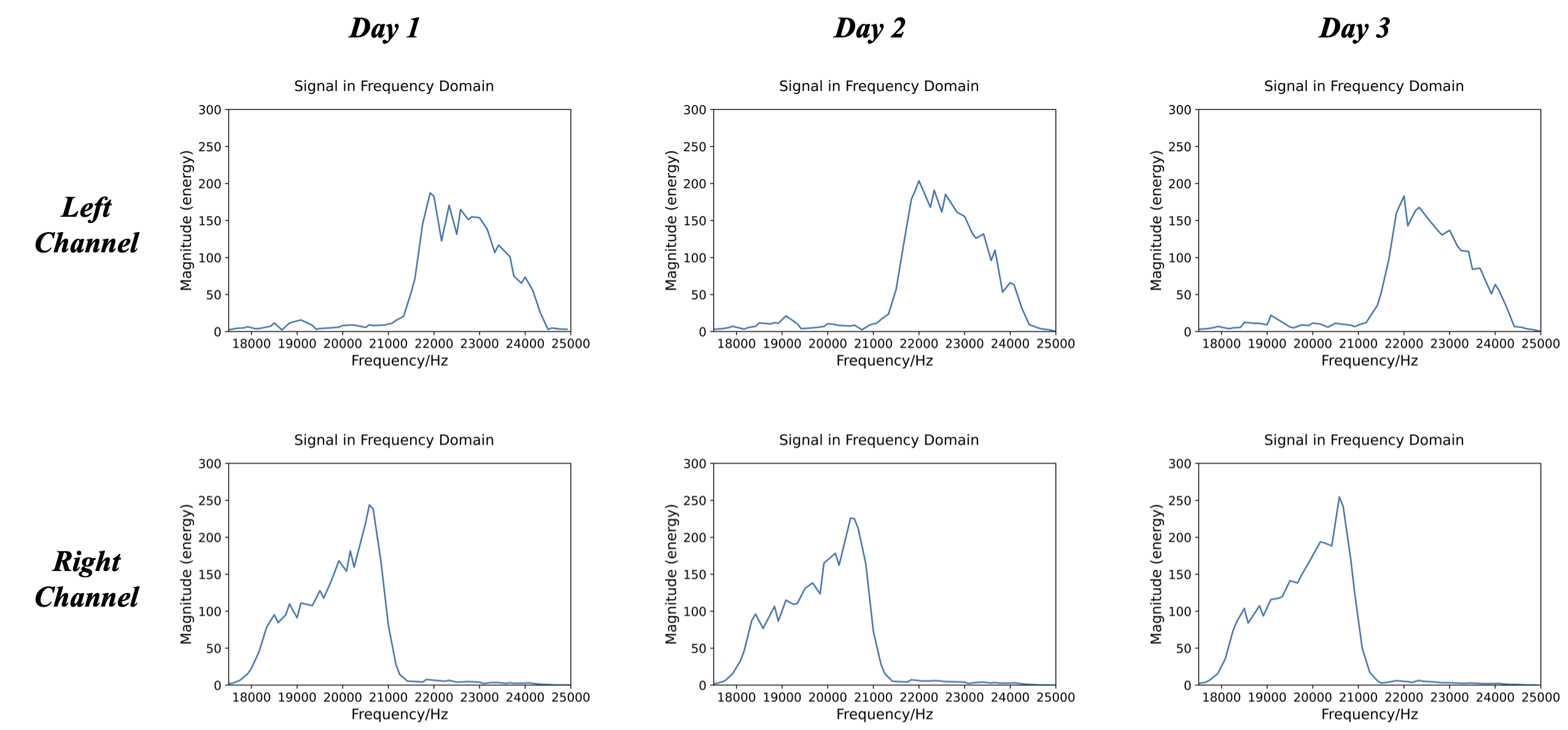}
    }\\
    
    \subfloat[Reflected Signals from P30]{
        \includegraphics[height=.45\textwidth]{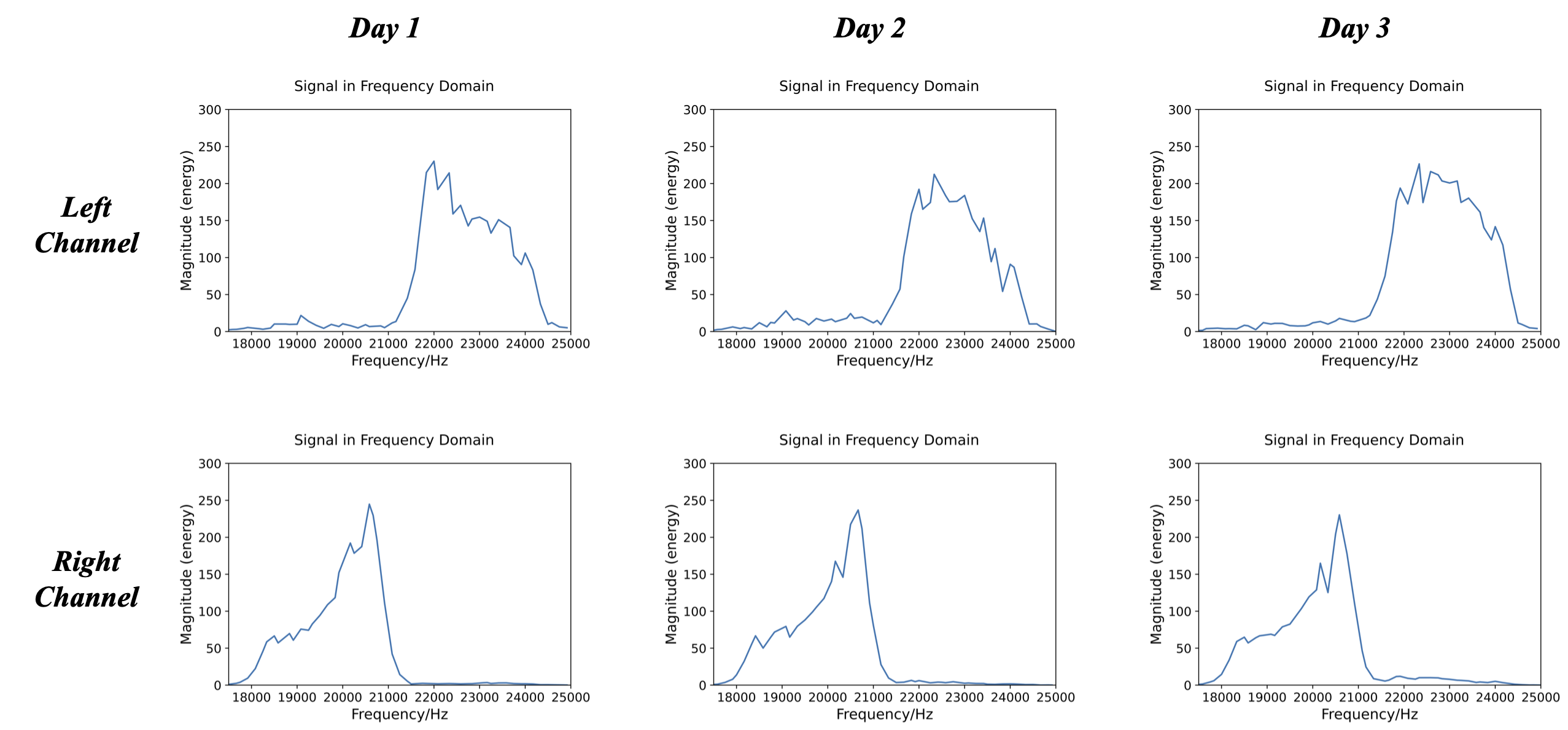}
    }
    \caption{Envelope of Reflected Signals in Frequency Domain from P25's Data and P30's Data across Three Days.}
    \label{Fig: frequency signal}
\end{figure}

We noticed that SonicID achieves promising performance at protecting the device from bad actors, even across several days. However, the system starts to experience more failure of authenticating the real users into the device across days, especially for several specific participants. This can be attributed to the lack of training samples from real users as we only trained the model with 1 minutes of data collected on the same day. However, the scanned acoustic features of users' faces can vary across days depending on their facial conditions such as how moist their faces are, the usage of makeups, the length of their beard, the coverage of their hair and so on. Usually the variation of the these factors should be subtle. However, under certain circumstances, these factors might significantly affect the reflection of the acoustic signals, leading to an unsatisfactory success rate of authentication. To validate our assumption, we plotted the reflected signals in the frequency domain from P25's data and P30's data on all three days. These two participants are selected because P30 has the lowest TPR on Day 3 and a low TPR on Day 2 (37.7\%) while P25 has a TPR of 100\% across all three days, if the model is trained on data from Day 1. To compare the change of signal reflection across days more clearly, we extracted the envelope of these signals to remove unnecessary details. The envelope signals are demonstrated in Fig.~\ref{Fig: frequency signal}.

As shown in the figure, there are subtle changes in the reflected signals across three days for P25 but overall they remain similar to each other. However, for P30, the signals largely change across different days. Specifically, signal reflection on Day 1 is quite different from the other two days. This explains why TPR of SonicID across days for P30 is worse than P25. As stated in Sec.~\ref{Sec: study results}, we believe that this may be owing to the long hair of P30 restraining the device from being put back to the same position every time of remounting. To address this issue, we decided to further fine-tune the individual model for participants with trouble logging into their device on a different day. The fine-tuning process of SonicID is that if one day a user figures that they need more than three trials to log into their device, i.e., the TPR of authentication on that day is lower than 33.3\%, they can activate the fine-tuning process in which the system collects 15 seconds of data from this user on that day to further fine-tune their trained individual model. This process can happen while the user is using the device after successfully logging into the system without interrupting their normal usage of the device. Generally, the fine-tuning process will only be carried out once so that limited burden of data collection will be added onto the user while the system guarantees a satisfactory authentication experience in the meantime.

Following this principle of fine-tuning, with the models trained on 1-minute data from Day 1, we fine-tuned the models for 4 participants on Day 2 (P32, P33, P36, and P40) and for 3 participants on Day 3 (P10, P27 and P30). The models were fine-tuned for 2 epochs. For the 4 participants with models fine-tuned on Day 2, the average TPR on Day 2 increases from 16.9\% to 74.3\% while FPR decreases from 4.6\% to 3.6\%. At the same time, the average TPR for these 4 participants on Day 3 also increases from 33.7\% to 57.7\% while FPR decreases from 4.8\% to 3.1\%. For the 3 participants with models fine-tuned on Day 3, the average TPR on Day 3 increases from 20.0\% to 63.3\% while FPR decreases from 6.7\% to 4.7\%. These results showcased that the performance of SonicID improves significantly in terms of both TPR and FPR across several days with only 15 seconds of data collected on a second day. Applying this fine-tuning process on the leave-one-day-out evaluation presented in Sec.~\ref{Sec: leave on day out}, we obtained the updated evaluation results in Tab.~\ref{Tab: different days with fine tune}. Tab.~\ref{Tab: different days with fine tune} displays that BAC is improved to around 90\% for the across day evaluation, with the fine-tuning process applied on only a few participants, which proved the feasibility of adopting fine-tuning to improve the performance of SonicID across days.

\begin{table}[t]
\caption{Average TPR, FPR and BAC across 40 Participants with Fine-tuning Process Applied on Selected Participants. Data Format: TPR/FPR/BAC. Results with Training and Testing Data from the Same Day are in Bold.}
\label{Tab: different days with fine tune}
\begin{tabular}{| c | c | c | c |} 
\hline
 Training Data Source & Day 1 & Day 2 & Day 3\\ [0.5ex] 
 \hline\hline
 Train on Day 1 & \textbf{97.4\%/4.3\%/96.6\%} & 86.5\%/4.0\%/91.3\% & 80.9\%/3.5\%/88.7\%\\
 \hline
 Train on Day 2 & 86.4\%/3.8\%/91.3\% & \textbf{96.5\%/3.6\%/96.5\%} & 84.1\%/3.7\%/90.2\%\\
 \hline
 Train on Day 3 & 81.3\%/3.5\%/88.9\% & 84.5\%/3.2\%/90.6\% & \textbf{98.2\%/3.1\%/97.5\%}\\
 \hline
\end{tabular}
\end{table}

\subsection{Impact of Different Body Postures}
\label{Sec: body postures}
As discussed in Sec.~\ref{Sec: user study}, the last 16 participants evaluated the system in various body postures including talking, standing and lying down. To explore the generalizability of SonicID when users are in different postures, we loaded the individual models trained on data collected when participants were sitting still and directly tested them with data collected when these participants were in different body postures. Note that although the positive samples only came from the data that each participant collected in specific scenarios, the negative samples came from the data all other participants collected in all scenarios, including sitting still. Same as Sec.~\ref{Sec: leave on day out}, leave-one-day-out evaluation was carried out to alleviate randomness across days. We averaged the results under two categories, i.e. performance on the same day of data collection and performance on a different day, and showcased the results in Tab.~\ref{Tab: different scenarios}.

\begin{table}[t]
\caption{Average TPR, FPR and BAC across Last 16 Participants for Different Application Scenarios. The Results are Averaged through Leave-One-Participant-Out Evaluation. Data Format: TPR/FPR/BAC.}
\label{Tab: different scenarios}
\begin{tabular}{| c | c | c | c |} 
\hline
 Scenarios & Same Day Evaluation & Across Day Evaluation & Across Day Evaluation with Fine-tuning\\ [0.5ex] 
 \hline\hline
 \textbf{Sitting Still} & \textbf{97.3\%/4.2\%/96.5\%} & \textbf{72.4\%/4.2\%/84.1\%} &
 \textbf{80.4\%/3.9\%/88.3\%}\\
 \hline
 Talking & 92.8\%/4.2\%/94.3\% & 68.3\%/4.2\%/82.0\% & 74.2\%/3.9\%/85.2\%\\
 \hline
 Standing & 92.2\%/4.2\%/94.0\% & 69.7\%/4.2\%/82.8\% & 73.9\%/3.9\%/85.0\%\\
 \hline
 Lying Down & 87.4\%/4.2\%/91.6\% & 62.9\%/4.2\%/79.3\% & 69.1\%/3.9\%/82.6\%\\
 \hline
\end{tabular}
\end{table}

As shown in Tab.~\ref{Tab: different scenarios}, in same day evaluation, TPR remains over 90\% for talking and standing scenarios and over 85\% for lying down scenarios even when the model is trained with data of sitting still only. BAC maintains over 90\% for all scenarios. In across day evaluation, FPR remains consistent under 5\% while TPR decreases compared with same day evaluation for all scenarios. BAC maintains around 80\% for all scenarios. Note that FPR for all scenarios under one evaluation is the same because the negative samples used to evaluate the model are the same. However, the negative samples used to evaluate the models in different evaluations are different. FPR remaining unchanged indicates that SonicID performs consistently well to reject unknown attackers across different days. With the fine-tuning process in Sec.~\ref{Sec: fine tune} applied, TPR across days improves by around 5\% and BAC is increased to around 85\% for all scenarios. Note that fine-tuning data only includes data collected when users are sitting still as well. These results proved that SonicID generalizes to various application scenarios well with only 1 minute of training data collected when the users are sitting. Specifically, we think that the selection of static instances contributed to the fact that the speech-related facial movements did not have a severe impact on the system performance because there were still relatively static moments, e.g. pauses between words, phrases and sentences, in each session even when the participants kept talking and the non-static data was not used for authentication in SonicID.

\begin{table}[t]
\caption{Average TPR, FPR and BAC across 40 Participants with Different Channels of Signals. Models are Trained with Data from Day 1. Data Format: TPR/FPR/BAC.}
\label{Tab: different sides}
\begin{tabular}{| c | c | c | c |} 
\hline
 Different Channels & Day 1 & Day 2 & Day 3\\ [0.5ex] 
 \hline\hline
 Both Sides & 97.4\%/4.3\%/96.6\% & 80.8\%/4.1\%/88.4\% & 75.2\%/3.8\%/85.7\%\\
 \hline
 Right Side & 90.2\%/3.6\%/93.3\% & 63.3\%/3.6\%/79.9\% & 52.7\%/3.6\%/74.5\%\\
 \hline
 Left Side & 86.0\%/5.8\%/90.1\% & 44.0\%/5.4\%/69.3\% & 31.1\%/4.7\%/63.2\%\\
 \hline
\end{tabular}
\end{table}

\subsection{Ablation Study}
\subsubsection{Comparison of Two Sides of Sensors}
When we designed the prototype for the SonicID system, we instrumented one pair of speaker and microphone on each side of the glasses to collect complete biometric information from users' face. However, many commercial smart glasses only have the speaker and microphone on one side of the device. As a result, we would like to experiment on how the system performs when only one side of sensors are utilized. As designed in Sec.~\ref{Sec: Design}, four speaker-microphone channels are input into the deep learning model when two pairs of speakers and microphones are deployed. If we only want to use one pair of speaker and microphone on either right side or left side of the glasses, we can just input one speaker-microphone channel related to this pair of speaker and microphone into the model. The experiment results are demonstrated in Tab.~\ref{Tab: different sides}. 

As we can see in the table, the system performance gets worse when only one side of sensors are used. We believe this is because deploying the sensors on one side only extract information from the channel of signal that travels on the same side of the face while instrumenting sensors on both sides also obtain information from the channel of signal that travels across users' face from one side to the other side. This variety of biometric information helps boost the system performance. In addition, one thing we noticed is that right side has much better performance than left side. Given people's face are mostly symmetric, we speculate that this discrepancy might be caused by the different frequency ranges we use on two sides (right side: 18-21 kHz and left side: 21.5-24.5 kHz). However, we believe that more thorough experiments are needed to verify this assumption. Even though one side of sensors cause the system performance to drop, the performance on Day 1 using right side sensors are still relatively comparable to using both sides sensors. This means that it is possible for SonicID to only include sensors on one side and more training data can be collected automatically when users are wearing the device to improve the system performance on following days.

\subsubsection{Impact of Pre-training Stage}
Sec.~\ref{Subsec: ML} introduced the two-stage machine learning pipeline we used to train the models and evaluate SonicID. In order to validate that the pre-training stage actually plays a positive role in the system, we run another experiment of directly training the feature extractor and binary classifier without the pre-training stage. All other parameters were kept the same and the models were trained on 1-minute data from Day 1 for each participant for 30 epochs in total. The average TPR/FPR/BAC for Day 1, Day 2 and Day 3 in this case are 96.6\%/4.0\%/96.3\%, 75.7\%/3.7\%/86.0\% and 68.4\%/3.3\%/82.6\% respectively while those for the original evaluation are 97.4\%/4.3\%/96.6\%, 80.8\%/4.1\%/88.4\% and 75.2\%/3.8\%/85.7\%. The comparison of the results displayed that the pre-training stage results in better BAC and TPR, especially on Day 2 and Day 3. This aligns with our intention of designing the pre-training stage in Sec.~\ref{Subsec: ML}, which is to better extract unique features associated with each participant that can help improve performance across days.

\subsubsection{Selection of Static Instances}
In the data processing pipeline of SonicID, the 10 most static 0.06-second instances are picked from each 10-second session to train the model and evaluate the system. These parameters were empirically selected in the pilot study. To showcase that the selection leads to reasonable performance of the system, we run experiments on using 5 and 15 most static instances to authenticate participants. Moreover, we experimented on different lengths of instances, including three 600-sample frames, i.e., 0.036 seconds of data and ten 600-sample frames, i.e., 0.12 seconds of data. The experimental results are displayed in Tab.~\ref{Tab: different instances}.

For all other selections of static instances except the combination we used in our evaluation, both TPR and FPR decreased on Day 2 and Day 3. In other words, the system was more strict in authenticating users. For the selection of 5 instances and 0.036-second instances, we believe that this was because only perfectly static instances were included in training and the model did not see enough variance so it could not be generalized to various scenarios well. For the selection 15 instances and 0.12-second instances, we assume that this was because the model was overfitted to the training data on Day 1 since the TPR on Day 1 was better but it failed to generalize across different days well. Despite of the variation among different selections of static instances, there was only slight divergence in terms of the balanced accuracy. As a result, the selection of ten 0.06-second static instances in our main study evaluation reached a trade-off balancing TPR and FPR.

\begin{table}[t]
\caption{Average TPR, FPR and BAC across 40 Participants with Different Selections of Static Instances. Models are Trained with Data from Day 1. Data Format: TPR/FPR/BAC. The Combination used in Main Study Evaluation is in Bold.}
\label{Tab: different instances}
\begin{tabular}{| c | c | c | c |} 
\hline
 Selection of Static Instances & Day 1 & Day 2 & Day 3\\ [0.5ex] 
 \hline\hline
 0.06 seconds, 5 Instances & 96.7\%/3.8\%/96.5\% & 75.7\%/3.5\%/86.1\% & 70.9\%/3.2\%/83.9\%\\
 \hline
 0.06 seconds, 15 Instances & 98.0\%/4.2\%/96.9\% & 78.6\%/4.0\%/87.3 & 71.1\%/3.5\%/83.8\%\\
 \hline
 \textbf{0.06 seconds, 10 Instances} & \textbf{97.4\%/4.3\%/96.6\%} & \textbf{80.8\%/4.1\%/88.4\%} & \textbf{75.2\%/3.8\%/85.7\%}\\
 \hline
 0.036 seconds, 10 Instances & 98.1\%/4.3\%/96.9\% & 79.7\%/4.2\%/87.8\% & 70.0\%/3.5\%/83.2\%\\
 \hline
 0.12 seconds, 10 Instances & 97.5\%/3.6\%/97.0\% & 78.1\%/3.4\%/87.3\% & 72.5\%/3.1\%/84.7\%\\
 \hline
\end{tabular}
\end{table}

\subsection{Usability}
\label{Subsec: usability}
A questionnaire was distributed to participants after the user study to collect their demographic information and feedback on the prototyped wearable device. First, the participants evaluated the overall experience of wearing the device by answering the question "How comfortable is this wearable device to wear around the face? (0 most uncomfortable, 5 most comfortable)". Among 40 participants, the average rating for this question is 3.4 (std=0.9), indicating that the device with the SonicID system is overall comfortable to wear around the face. Next we specifically asked participants to evaluate the weight of the device with the question "How acceptable do you find the weight of our wearable device? (0 most unacceptable, 5 most acceptable)". The average rating is 3.8 (std=0.9), validating that the weight of the device does not cause much burden on users. 

In the SonicID system, we utilized two ultrasonic signals in the frequency ranges over 18 kHz. Theoretically, most people are not capable of hearing the sounds emitted from the system. However, due to the limitation of the hardware components, there might be some frequency leakage into the lower frequency range and some users might be able to hear the sound. As a result, we asked the participants to also answer the question "Can you hear the sound emitted from our system?". 29 out of 40 participants answered "No" to this question while the other 11 participants reported "Yes". For those participants who answered "Yes" to this question, a follow-up question "If yes, how comfortable do you feel about the emitted sound? (0 most uncomfortable, 5 most comfortable)" was asked. The average rating for this question among 11 participants is 4.4 (std=0.5). Even though some users might hear the sound from the system, they generally find this sound not bothering them a lot. Since the echo profile we use as an instance to authenticate users is as short as 0.06 s as discussed in Sec.~\ref{Sec: echo profile calculation}, we do not expect the sound to trouble the users continuously. Furthermore, it is possible to lower the strength of the transmitted signals to make them less noticeable by users.

Finally, the participants provided some open-ended comments on the wearable device we prototyped. Most participants found this device easy to wear and no discomfort while wearing it. Some participants reported that the glasses was a little imbalanced due to the placement of the micro-controller. Some participants stated that their hair was taken away by the micro-controller when they taken off the device. Other participants believed that the placement of the micro-controller and the wires can be improved to make the prototype more comfortable. These issues can be addressed by balancing the two sides of the glasses and better incorporating the micro-controller into the glasses with some cases or covers. Besides, some participants thought that the glasses was a bit tight for them. This is the problem of the commodity glasses we use instead of the SonicID system itself. However, we can potentially implement several prototypes of the system on glasses of different sizes to tackle this problem.

%% file: 8_discussion.tex
\section{Discussion}
\label{Sec: Discussion}

\subsection{Power Consumption}
We wanted to evaluate the power consumption of the SonicID system in operation. For this purpose, we measured the current that flowed through system with a current ranger\footnote{\url{https://lowpowerlab.com/guide/currentranger/}} when the system was operating. The current was measured at 165.4 mA with a voltage at 3.3 V. This gave us an average power consumption of 545.8 mW. Considering that SonicID only needs an instance of 0.06 s to authenticate users every time, theoretically it consumes 32.7 mJ of energy for each authentication trial. Note that the calculation of the power consumption here does not include that of the machine learning inference because the inference stage is currently carried out on a local server. To implement a real-time pipeline for SonicID, we could run the model inference stage on a low-power chip powered with a CNN accelerator, such as MAX78002~\cite{MAX78002}. Prior work~\cite{li2024gazetrak} presented that ResNet-18 can be run on MAX78002 to make 30 inferences per second with a power consumption of 79.0 mW. Hence, we anticipate that the machine learning pipeline of SonicID can also be deployed on MAX78002 to realize a real-time pipeline since we also use a model based on the ResNet-18 architecture. Under this circumstance, SonicID needs 0.093 s in time and 58.1 mJ in energy to perform each trial of user identification. If SonicID is deployed on Meta Ray-Ban with a battery capacity of 154 mAh, the authentication system can be activated around 31,500 times in theory, if SonicID is used alone. However, we believe that these metrics need to be measured more accurately when a real-time pipeline is actually implemented. We leave this for future work.

\subsection{Health Implications}
Even though we picked the signals in the ultrasonic frequency ranges to make them inaudible to the users, they may still cause health concerns because the system is close to users' ears when they wear it. Therefore, we used the NIOSH sound level meter App\footnote{\url{https://www.cdc.gov/niosh/topics/noise/app.html}} to measure the sound level of the signals emitted from the system. We kept the system transmitting signals continuously and placed the microphones of the smartphone with the App running at a distance, where users' ears would be if they wore the glasses, from the speakers of the SonicID system. The sound level was measured at 65.8 dB. Based on the findings in \cite{howard2005review}, the recommended exposure limit for ultrasonic signals around 20 kHz is 75 dB. This indicates that our SonicID system is safe to wear for the general public given that the authentication process only lasts for a short period of time when users start to operate the device. What is more, the signal strength can be reduced to further restrain its impact on users' health.

\subsection{Applications}
SonicID proposes a low-power and minimally-obtrusive solution to user authentication on smart glasses. Meanwhile, the current prototype is relatively cheap (around \$50) with a lot of potentials for further modification and optimization with an even cheaper cost. Therefore, we expect the system to be applied on more wearable devices such as VR headsets, AR glasses, etc. The authentication pipeline can also be utilized in scenarios beyond that when users trying to log into the device. It can be used anytime when you need user identification on wearable devices, including making payments, logging into a social account, changing the settings of the device, etc. Despite the promising application space of SonicID, it is important to keep improving its performance to make sure that the false positive samples are as few as possible.

\subsection{Potential Methods to Improve Performance}
\label{Subsec: improve performance}
As evaluated in Sec.~\ref{Sec: Evaluation}, although the FPR is consistently below 5\% across several days in evaluation, the TPR reduces on Day 2 and Day 3 compared with Day 1 for some participants. We believe this is due to several reasons. Firstly, the positive samples we collected from each participant every day were just from the 5 minutes of data. The small size of the data could lead to fluctuations in the evaluation results in TPR. Secondly, we noticed that TPR is especially low for P30 and P32 who have long hair and bangs. The hair could sometimes cover the system and prevent the participant from putting the glasses back to the same position on their head every time they remounted the device. We believe that the same reason caused P4 and P27 to have a worse FPR than other participants. In future, we plan to explore several potential methods to improve the performance of the system. 

\subsubsection{Explore Movement-based Authentication}
In Sec.~\ref{Sec: Evaluation}, we removed information of blinking and other movements from the data during training and testing to guarantee a valid facial scanning outcome. However, this information can also be taken advantage of to authenticate users. For instance, the blinking information alone can be exploited as a unique biometric pattern because different users blink differently. Furthermore, we can explore other movement-based authentication methods to add another layer of protection. Users of our system can select a specific facial gesture (e.g. smile, tongue out, blink, etc.) as their password to the system. Our system recognizes both their password and their biometric information when performing this password to authenticate users. This kind of two-layer authentication systems make the system more secure and are possible to implement given that there are already some work validating that deploying acoustic sensing on wearables can help track users' facial movements~\cite{li2022eario,zhang2023i,xie2021acoustic,li2024eyeecho}. SonicID can potentially be integrated into these existing systems to realize user authentication without adding extra hardware.

\subsubsection{Continuously Collect Training Data}
\label{Subsubsec: more data}
Another potential way to improve the system performance is to continuously collect training data while the user operates the glasses. Currently, SonicID runs a registering process for new users to collect training data. This small amount of training data limits the stability of the system across different days due to the lack of variance, especially for TPR. SonicID can alleviate this problem by continuously collecting the biometric information from the user's face while they use the smart glasses. The system can scan their face occasionally when they use the device and pick the static instances to train and fine-tune the model. This helps reduce the burden of collecting too much training data before using the authentication system. The fine-tuning experiments in Sec.~\ref{Sec: fine tune} proves the feasibility of this method.

\subsubsection{Pre-collect More Data in Pilot Study}
While we trained the models in the user study, the negative samples came from the 21 people from whom we collected data in the pilot study. If we can recruit more volunteers to collect negative samples in this dataset, the system performance, especially FPR, can be improved since the model will learn the acoustic patterns better from more data. In the meantime, this does not add extra burden of collecting more training data on the new users.

\subsection{Limitation and Future Work}
\label{Subsec: limitation} 
SonicID provides a low-power, effortless, and minimally-obtrusive solution to user authentication on smart glasses that works relatively consistently well across different remounting session and different days. However, there are still some limitation of this work.

\subsubsection{Impact of Long Hair}
The long hair and bangs of users might impact the performance of SonicID and even cover the system. In the study, participants with long hair were asked to put their hair up with a hair tie. Nevertheless, some participants experienced difficulty putting the glasses back to the same wearing position every time and had worse performance in TPR or FPR, including P4, P27, P30 and P32 that have been discussed in Sec.~\ref{Sec: study results}, compared with other participants. In the future, we plan to further upgrade the prototype by placing the MCU and sensors at a better position and covering them with a case so that they are not easily impacted by long hair and bangs.

\subsubsection{Amount of Training Data Needed}
Sec.~\ref{Sec: training data amount} evaluates the performance of SonicID with different amounts of training data. With 1 minute of data, SonicID achieves satisfactory performance on Day 1 but the performance, especially TPR decreases on Day 2 and Day 3. One solution to this problem is to collect more training data when the user operates the device continuously in the first couple days of wearing the device as discussed in Sec.~\ref{Subsubsec: more data}. Moreover, when we are able to collect data from more participants in the future, the model can better learn the acoustic patterns and make better authentication.

\subsubsection{Adapt to Commodity Smart Glasses} 
To validate the feasibility of SonicID, we deployed the system on a common pair of glasses and conducted the study. In future, we plan to directly use the built-in speakers and microphones on commercial smart glasses to implement the system so that it is directly applicable and ready to use by customers.

%% file: 10_conclusion.tex
\section{Conclusion}
\label{Sec: Conclusion}
In this paper, we present an acoustic-based solution to user identification on smart glasses, called SonicID. It adopts active acoustic sensing to scan the user's face so that biometric information is obtained from the user to authenticate them. A customized binary classifier based on the ResNet-18 architecture is used to extract unique features related to each user from the acoustic patterns. A user study with 40 participants validated the performance of SonicID across different remounting sessions and days. SonicID authenticates the user by scanning their face for 0.06 seconds as soon as they put on the device and only consumes 32.7 mJ of energy for each authentication trial, excluding the machine learning model inference part. We further explore SonicID's performance under different settings and discuss the potential applications of SonicID and how it can be deployed on commercial smart glasses in the future.

%% file: 11_acks.tex
\begin{acks}
This work is supported by National Science Foundation (NSF) under Grant No. 2239569, NSF's Innovation Corps (I-Corps) under Grant No. 2346817, the IGNITE Innovation Acceleration Program, and the Ann S. Bowers College of Computing and Information Science at Cornell University. This paper includes sections refined with the assistance of ChatGPT. 
\end{acks}

%% file: 12_appendix.tex
\appendix
\newpage
\section*{APPENDIX}
\section{Individual Evaluation Results}\label{Appendix: individual results}

\begin{figure*}[htbp]
  \includegraphics[width=0.81 \textwidth]{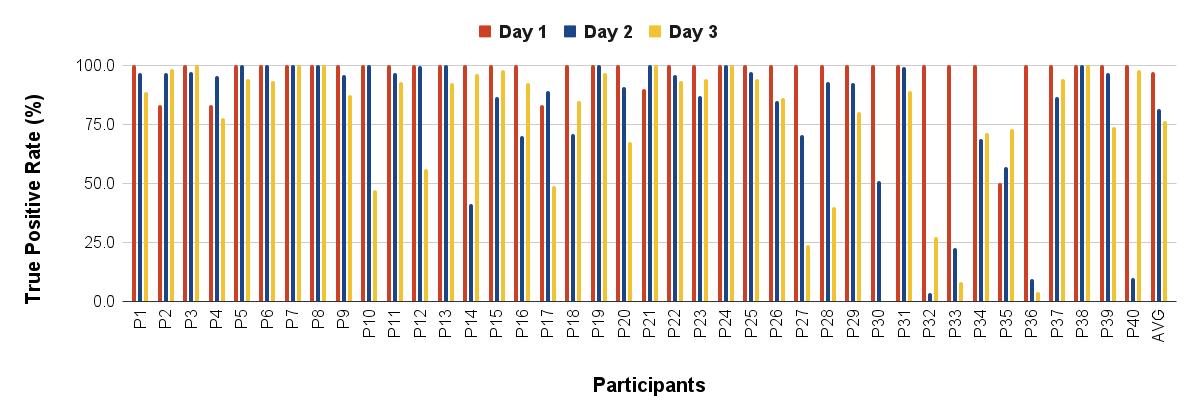}
  \caption{True Positive Rate for all Participants across Three Days.}
  \label{Fig: results tpr}
\end{figure*}

\begin{figure*}[htbp]
  \includegraphics[width=0.81 \textwidth]{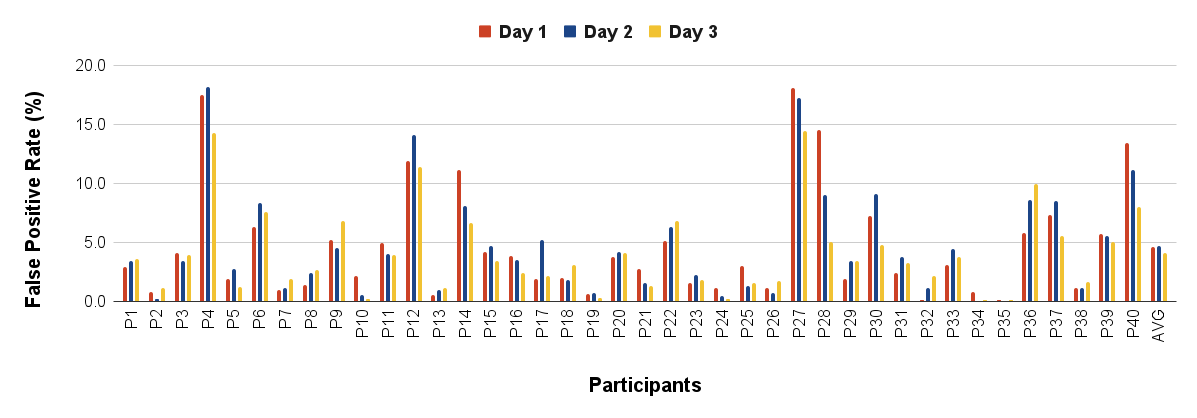}
  \caption{False Positive Rate for all Participants across Three Days.}
  \label{Fig: results fpr}
\end{figure*}

\begin{figure*}[htbp]
  \includegraphics[width=0.81 \textwidth]{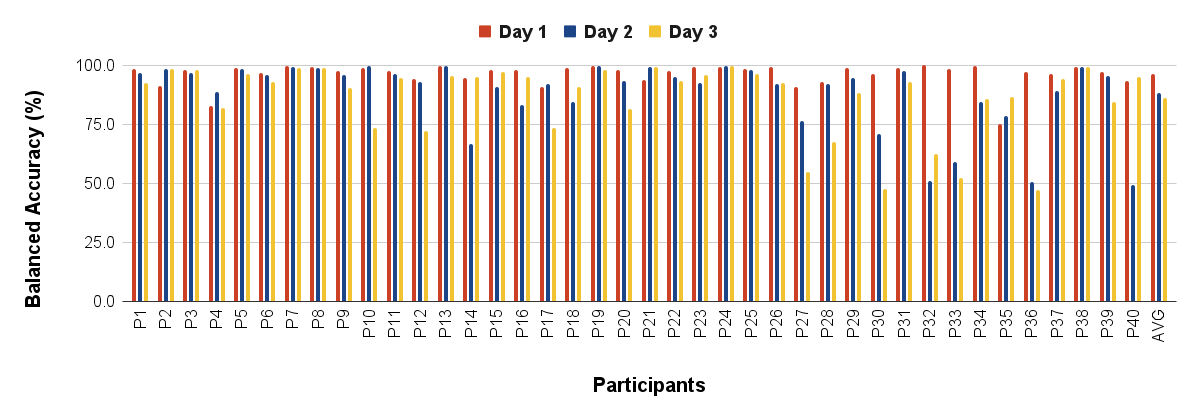}
  \caption{Balanced Accuracy for all Participants across Three Days.}
  \label{Fig: results ba}
\end{figure*}